\documentclass{aa}
\usepackage{graphicx}
\usepackage{color}
\usepackage{amsbsy}
\usepackage{mathrsfs}
\usepackage{mathpazo,bm}
\usepackage{amsmath}
\usepackage{multirow}
\usepackage{esint}

\newcommand{\pderiv}[2]{\frac{\partial #1}{\partial #2}}
\newcommand{\pderivn}[3]{\frac{{\partial{}}^{#3} #1}{{\partial #2}^{#3}}}
\newcommand{\aderiv}[1]{\frac{D #1}{Dt}}
\newcommand{\ee}{\mathrm{e}}
\newcommand{\ttimes}[1]{10^{#1}}
\newcommand{\xtimes}[2]{#1 \times 10^{#2}}

\newcommand{\vv}[1]{{\boldsymbol #1}} 

\newcommand{\del}{\vv{\nabla}}
\newcommand{\Div}{\vv{\nabla}\cdot}
\newcommand{\curl}{\vv\nabla\times}
\newcommand{\Laplace}{\nabla^2}
\newcommand{\de}{\mathrm{d}} 
\newcommand{\rint}{$s_{\rm int}\,$}
\newcommand{\rext}{$s_{\rm ext}\,$}
\newcommand{\hats}{\hat{\vv{s}}}
\newcommand{\hatphi}{\hat{\vv{\phi}}}
\newcommand{\hatz}{\hat{\vv{z}}}
\newcommand{\va}{v_{_{\rm A}}}

\newcommand{\mm}[2]{#1{\tiny$\pm$#2}}

\newcommand{\App}[1]{Appendix~\ref{#1}} 

\begin{document}

\title{Global magnetohydrodynamical models of turbulence in protoplanetary
disks}
\subtitle{I. A cylindrical potential on a Cartesian grid and transport of solids}

\author{W. Lyra\inst{1},~
  A. Johansen\inst{2},~
  H. Klahr\inst{2},~and
  N. Piskunov\inst{1}}

\offprints{wlyra@astro.uu.se}

\institute{Department of Astronomy and Space Physics, Uppsala Astronomical Observatory, Box 515, 751\,20 Uppsala, Sweden
\and Max-Planck-Institut f\"ur Astronomie, K\"onigstuhl 17, 69117 Heidelberg, Germany 
}

\date{Received ; Accepted}

\authorrunning{Lyra et al.}
\titlerunning{Disk-in-a-box}

\abstract
{}
{We present global 3D MHD simulations of disks of gas and solids, aiming at
developing models that can be used to study various scenarios of planet
formation and planet-disk interaction in turbulent accretion disks.} 
{We employ the {\sc Pencil Code}, a 3D high-order finite-difference MHD code
using Cartesian coordinates. We solve the equations of ideal MHD with a local
isothermal equation of state. Planets and stars are treated as particles
evolved with an $N$-body scheme. Solid boulders are treated as individual
superparticles that couple to the gas through a drag force that is linear in
the local relative velocity between gas and particle.}
{We find that Cartesian grids are well-suited for accretion disk problems. The
disk-in-a-box models based on Cartesian grids presented here develop and
sustain MHD turbulence, in good agreement with published results achieved with
cylindrical codes.We investigate the dependence of the magnetorotational
instability on disk scale height, finding evidence that the turbulence
generated by the magnetorotational instability grows with thermal pressure. The
turbulent stresses depend on the thermal pressure obeying a power law of
$0.24\pm0.03$, compatible with the value of $0.25$ found in shearing box
calculations. 
The ratio of Maxwell to Reynolds stresses decreases with increasing 
temperature, dropping from 5 to 1 when the sound speed was raised by a 
factor 4, maintaing the same field strength.
We also study the dynamics of solid boulders in the hydromagnetic turbulence, 
by making use of $10^6$ Lagrangian particles embedded in the Eulerian grid. 
The effective diffusion provided by the turbulence prevents settling of the
solids in a infinitesimally thin layer, forming instead a layer of solids of 
finite vertical thickness. The measured scale height of this 
diffusion-supported layer of solids implies turbulent vertical diffusion 
coefficients with globally averaged Schmidt numbers of
1.0$\pm$0.2 for a model with $\alpha\approx\ttimes{-3}$ and 0.78$\pm$0.06 for a
model with $\alpha\approx\ttimes{-1}$. That is, the vertical turbulent 
diffusion acting on the solids phase is comparable to the turbulent 
viscosity acting on the gas phase. The average bulk density of solids in the 
turbulent flow is quite low 
($\rho_{\rm p}$=$\xtimes{6.0}{-11}\,\rm{kg\,m^{-3}}$), but in the high 
pressure regions, significant overdensities are observed, where the 
solid-to-gas ratio reached values as great as 85, corresponding to 4 
orders of magnitude higher than the initial interstellar value of 0.01}
{} 

\maketitle

\section{Introduction}

Planets have long been believed to form in disks of gas and dust around young
stars (Kant 1755, Laplace 1796), interacting with their surroundings via a set
of complex and highly nonlinear processes. In the core accretion scenario for
giant planet formation (Mizuno 1980), dust coagulates first into km-sized icy
and rocky planetesimals (Safronov 1969, Goldreich \& Ward 1973, Youdin \& Shu
2002) that further collide, forming
progressively larger solid bodies that eventually give rise to cores of several
Earth masses. If a critical mass is attained, these cores become gas giant
planets by undergoing runaway accretion of gas (Pollack et al. 1996).
Otherwise, just a small amount of nebular gas is retained by the core, which
ends up as an ice giant. 

The success of this picture in explaining the overall shape of the solar system
was shaken by the discovery of the extra-solar planets. In less than a decade,
the zoo of planetary objects received exotic members such as close-in Hot
Jupiters (Mayor \& Queloz 1995), pulsar planets (Wolszczan \& Frail 1992),
highly eccentric giants (Marcy \& Butler 1996), free-floating planets (Lucas \&
Roche 2000), and super-Earths (Rivera et al. 2005). Thus, understanding the
diversity of these extra-solar planets is a crucial task in planet
formation theory.

Planet-disk interaction seems to be one of the obvious candidates to account
for this diversity. Planets exchange angular momentum with the disk,
leading to either inward or outward migration (Ward 1981; Lin and Papaloizou
1986; Ward \& Hourigan 1989; Masset et al. 2006). An understanding of the
physical state of accretion disks is essential to provide a detailed picture of
the effect of migration on planetary orbits.

Analytical theory must necessarily contain a number of linearizing
simplifications. Therefore, numerical simulations are a major tool to provide
advances in the problem. But even then, the large computational demands of such
calculations have put some restrictions and limitations in the models presented
so far. Because of this, although many of the individual physical processes
occurring on circumstellar environments are understood in some detail,
state-of-the-art calculations on planet formation still lag behind our current
understanding, containing simplifying assumptions needed to reduce the
computational effort.

For example, the evolution of temperature is usually neglected in solving the
dynamical equations, favoring an imposed temperature profile. Paardekooper \&
Mellema (2006) showed that in non-isothermal disks, the net torques acting on a
forming planet can change sign due to asymmetric heating on the planet's 
corotation region, potentially stopping and reversing the migration of the
planet. 2D and 3D models of disks with radiative transfer were presented by
D'Angelo et al. (2003) and Klahr \& Kley (2006), showing that a high-mass
planet may carve a cold gap in the disk while retaining a thick circumplanetary
cloud. But no radiative global simulation with explicit ray tracing,
able to consistently treat optically thin and thick regions and the transition
between them, has been presented so far.

Magnetic fields have been shown to play a major role in the structure
and evolution
of accretion disks. Observational efforts in the detection and analysis of
protoplanetary disks show evidence that these disks accrete, with a mass
accretion rate of the order $\approx \ttimes{-8} {\rm M_{\sun} yr^{-1}}$ (e.g.,
Sicilia-Aguilar et al. 2004). Such a powerful accretion cannot be explained by
molecular viscosity, requiring some other mechanism to transport angular
momentum outward. Balbus \& Hawley (1991) pointed out the importance of the
magnetorotational instability (Velikhov 1959, Chandrasekhar 1960, 1961) for
accretion disks. In their important work, they show that this magnetorotational
instability (MRI) is operative in sufficiently ionized Keplerian disks as long
as the magnetic field is subthermal, generating a turbulence powerful enough to
explain observed accretion rates in protoplanetary disks.

However, although magnetic fields are ubiquitous in the universe,
protoplanetary disks are thought to be ``cold'' and thus not completely
ionized. Cosmic rays can provide the required ionization for the MRI to
operate, but they cannot penetrate all the way to the midplane of the disk (a
standard value for the penetration depth is a gas column density of
$\Sigma=100\,{\rm g\,cm^{-2}}$). The result is that in the region 
where giant planets
are thought to form, only the surface of the disk is sufficiently ionized for
the MRI to grow. Turbulence thus likely operates in a surface layer, while the
midplane is neutral and laminar, constituting a so called ``dead zone'' (Gammie
1996, Miller \& Stone 2000, Oishi et al. 2007).

As a result of the mentioned difficulties of modeling the coupled interaction
between radiation, magnetic fields, dust grains, solids, neutral and ionized
gas in the gravitational potential of a star and embedded planets, the
numerical works in the field show a heterogeneity of methods, with most works
tackling only some aspects of the problem. Particularly, numerical simulations
have focused on local Cartesian shearing boxes (e.g., Hawley et al. 1995,
Brandenburg et al. 1995) or global disks on cylindrical grids (e.g., Hawley
2001, Armitage et al. 2001, Nelson 2005). As the MRI is a local process, the
shearing box has the advantage of capturing much of the physics of the problem
while significantly reducing the computational effort and complexity --  for instance (shear-)
periodic boundary conditions can be used. The global disks on cylindrical grids
offer the advantage of having the grid and flow geometry coinciding, but
in this case, special care must be taken for the boundary conditions, as
reflective boundaries make waves bounce through the computational domain in an
unphysical manner and outflow boundary conditions may lead to too much mass
loss (Fromang \& Nelson, 2006).  Fromang \& Nelson (2006) have also presented
the first simulation of the MRI in global disks with vertical density
stratification. A comparison between their models and a stratified version of
ours will be addressed in future work.

In a series of articles we aim at constructing global radiative
magnetohydrodynamical simulations. In this first paper, we present the features
and capabilities of the numerical scheme used by constructing cylindrical disk
models of gas and solids with MHD turbulence.  These models will be 
developed in future work to allow for stratification, radiation and a global
self-consistent treatment of dead zones.

The simulations presented here are embedded in Cartesian boxes. Although it can
be regarded as unpractical for simulating a flow with cylindrical symmetry,
such a grid also presents  some advantages.  First, cylindrical grids are a
strong limitation for flows that deviate from cylindrical symmetry, e.g.\ 
circumbinary disks or 3D jet simulations, mainly because it is impossible to
have a flow across the center of the grid, and at $r$=0 reflection must occur.
Second, this approach has proved useful in view of computational simplicity and
parallelization efficiency (e.g.\ Dobler et al.  2006). In particular, by having
cells of constant aspect ratio, Cartesian grids have much reduced numerical
dissipation when compared to grids with complex geometry (van Noort et al.
2002). Therefore, while cylindrical and spherical grids can explicitly conserve
angular momentum w.r.t the origin of the coordinate system, it is of no benefit
for systems that do not have the center of mass at the origin. Third, as
photons travel in straight lines (in the absence of general relativistic
effects), a radiative transfer scheme with ray tracing is simpler to implement
in a Cartesian grid in spite of the cylindrical symmetry of the hydrodynamical
flow (Freytag et al. 2002). 

As a numerical solver we employ the {\sc Pencil Code}{\footnote{See
http://www.nordita.dk/software/pencil-code}}, a high (6th) order finite difference code.
Such a numerical tool is highly different from most other astrophysical
codes in use in the literature (see de Val-Borro et al. 2006 and references
therein), thus also providing an independent check of the results so far
obtained in the field. 

This paper is structured as follows: we discuss the model in Sect. 2,
proceeding to test cases in Sect. 3. In Sect. 4 we discuss the several MHD
simulations performed.  In Sect. 5 the models with solids are presented,
finally leading to the conclusions in Sect. 6.

\section{The model}
\subsection{Gas dynamics}

The equations solved are those of ideal MHD in an inertial reference frame with
a central gravity source. The equation governing the evolution of density is
the continuity equation
\begin{equation}
\label{continuity}
  \aderiv{\rho} = -\rho{\Div\vv{u}} + f_{_D}(\rho),
\end{equation}
where $\rho$ and $\vv{u}$ are the density and velocity of the gas.  The operator
${D}/Dt = \partial /{\partial}t + \vv{u}\cdot\del$ represents the advective
derivative.

The equation of motion is the sum of all forces acting on a parcel of gas. It
reads
\begin{equation}
\label{navierstokes}
  \aderiv{\vv{u}} = -\frac{1}{\rho}\del{p} - 
\del\Phi + \frac{\vv{J}\times\vv{B}}{\rho} + \vv{f}_{\nu}(\vv{u},\rho),
\end{equation}
where $p$ is pressure, $\Phi$ the gravitational potential, $\vv{B}$ is the
magnetic field, $\vv{J}=\mu_0^{-1}\vv{\nabla}\times\vv{B}$ is the volume current
density, as defined by Amp{\`e}re's Law, and $\mu_0$ is the magnetic
permeability of vacuum. 

The evolution of the magnetic field is governed by the induction equation. The
{\sc Pencil code}, however, works not with the magnetic field itself, but with
the magnetic potential $\vv{A}$, where $\vv{B}=\curl\vv{A}$. This automatically
guarantees the solenoidality of the magnetic field, as the condition $\Div\vv{B}
= \Div(\curl\vv{A})=0$ is always satisfied. The induction equation formulated
for the magnetic potential reads
\begin{equation}
\label{induction}
  \pderiv{\vv{A}}{t} = \vv{u}\times\vv{B} + \vv{f}_{\eta}(\vv{A}).
\end{equation}
The equation of state, relating pressure and density, closes the system of
equations. We use the ideal gas law 
\begin{equation}
\label{pressure}
   p=\rho c_s^2,
\end{equation}
with a locally isothermal approximation, where the sound speed $c_s$ is a
time-independent function of the cylindrical distance $s$ to the $z$-axis. We
write cylindrical coordinates as ($s$,$\phi$,$z$) and spherical coordinates as
($r$,$\phi$,$\theta$), where $\theta$ is the polar angle and $\phi$ the
azimuthal angle. The $z$ direction is perpendicular to the midplane of the disk.

The gravitational potential $\Phi$ has contributions from the star and the
$N-1$ embedded planets,
\begin{equation}
\label{gravitational-potential}
  \Phi = -\sum_{i}^{N}{\frac{GM_i}{\sqrt{{\mathcal R}_i^2+b_i^2}}}, 
\end{equation}
where $G$ is the gravitational constant, $M_i$ is the mass of particle $i$ and
${\mathcal R}_i=|\vv{r}-{\vv{r}}_{p_i}|$ is the distance of a gas parcel relative
to particle $i$. The quantity $b_i$ is the distance over which the gravity
field of the particle $i$ is softened to prevent singularities.

The functions $f_{_D}(\rho)$, $\vv{f}_{\nu}(\vv{u},\rho)$, and
$\vv{f}_{\eta}(\vv{A})$ are explicit mass diffusion, viscosity and
resistivity terms,
needed to stabilize the numerical scheme. They are composed of two terms, where
the first one is a conservative sixth-order dissipation. This term is described
in detail in Haugen \& Brandenburg (2004) as well as in Johansen \& Klahr
(2005) for the case of isotropic dissipation. A generalization for the
anisotropic case, required for non-cubic cells, is shown in
\App{ch:hyperdissipation}. The second term is a localized shock-capturing
dissipation, activated when large negative divergences, typical of shocks, are formed
(Haugen et al. 2004). This is described in \App{ch:shocks}.
 
\subsection{Planet orbital evolution}

The star and the planets are treated as an $N$-body ensemble, evolving
due to
their mutual gravitational interaction. The equation of motion for particle $i$
is 
\begin{equation}
\label{dust-dynamics}
  \frac{d{\vv{v}}_{p_i}}{dt} = {\vv{F}}_{g_i} -\sum_{j\neq i}^{N}\frac{G M_j}{{\mathcal R}_{ij}^2} \hat{\vv{{\mathcal R}}}_{ij}
\end{equation}
where ${\mathcal R}_{ij}=|\vv{r}_{p_i}-{\vv{r}}_{p_j}|$ is the distance between
particles $i$ and $j$, and $\hat{\vv{{\mathcal R}}}_{ij}$ is the unit vector
pointing from particle $j$ to particle $i$. The first term of the R.H.S.\ is the
combined gravity of the gas onto the particle $i$
\begin{equation}
\label{disk-gravity}
  {\vv{F}}_{g_i} = -G\int_{V} \frac{\rho(\vv{r})\vv{{\mathcal R}}_i}{({\mathcal R}_i^2 + b_i^2)^{3/2}} \de V, 
\end{equation}
where the integration is carried out over the whole disk. As we are not
interested in the disk's self-gravity, but rather on its gravitational effect
on one specific point (or a few points in case of multiple planets),
calculating the integral above is simpler and faster than using a Poisson
solver to find the gravitational potential of the disk everywhere on
the grid.

The smoothing distance $b_i$ is taken to be as small as possible. It is usually
a fraction of the Hill radius. For reasons described in sect. 2.7, the stellar
potential can be treated as unsoftened ($b_\star$=0). We note that in this
formulation there is no distinction between a planet and a star except for the
mass. The star evolves dynamically due to the gravity of the planets, wobbling
around the center of mass of the system, which is set to the center of the
grid. As the disk is not massive compared to the star, we exclude the disk
torques from influencing the star. For runs without planets a constant gravity
profile with a star at the center of the grid is used instead of solving the
equations of the $N$-body code. 

\subsection{Dynamics of solids}

To model the early stages of planet formation where solids grow from cm
and m sizes to kilometer-sized planetesimals we consider the dynamics
of meter-sized solid boulders, also treated as individual Lagrangian
particles. Each of the particles has its own position and velocity,
independent of the grid, integrated by the same particle module of the {\sc
Pencil Code} that is used for
the planets. The difference is that as the planets interact with the disk and
with themselves by gravity, the particles interact with the disk only via a
drag force that is proportional to the velocity of the particle with respect to
the local gas velocity. 

While in our cylindrical models there is no vertical gravity on the gas, the
particles do feel this component without which no settling towards the disk
midplane would occur. The evolution equation for solid particle $i$ is
therefore
\begin{equation}
  \frac{d{\vv{v}}_{i}}{dt} = -\frac{1}{\tau_f}(\vv{v}_{i}-\vv{u}) -\frac{GM_\star}{r^3}\vv{r},
\end{equation}
where $\tau_f$ is the friction time and $\vv{u}$ is the gas velocity at the
position of a particle. We assume that the friction time is independent of
velocity differences between gas and particles. We choose it to be
$\tau$=1/$\Omega_0$, which for the typical densities and temperatures in
the disk (Table 1), corresponds to particle radii between 0.4 and 2.5 meters,
depending inversely on the orbital distance. The assumption of linearity of 
the drag law holds as long as the velocity difference between gas and 
solids is much smaller than the sound speed (Weidenschilling 1977, 
Paardekooper 2007, Johansen et al. 2007). The condition is met since the 
turbulence generated by the MRI is subsonic.

The gas velocity $\vv{u}$ at the position of the particle is interpolated from
the nearest 27 grid points, using a Triangular Shaped Cloud scheme, as
described in Youdin \& Johansen (2007).

\subsection{The code}

The {\sc Pencil code} is a non-conservative Cartesian
finite-difference MHD code that uses sixth order centered spatial derivatives
and a third order Runge-Kutta time-stepping scheme, being primarily designed
for compressible turbulent hydromagnetic flows. 	

The {\sc Pencil code} was recently applied to a 2D global laminar disk
calculation, in which the results agreed with those of polar-grid based codes
(de Val-Borro et al. 2006). We extend this calculation now to three dimensions
with magnetic fields, fully exploiting the capabilities of the {\sc Pencil
code} for handling the problem of numerical hydromagnetic turbulence.   

\subsection{Units}

We adopt dimensionless units such that \[GM=\rho_0=\mu_0=1\] The quantity $GM$
has dimension of $\rm{length^3\,time^{-2}}$, so it sets a constraint on
[$x$][$t$]. The unit of time follows from this as being the inverse of the 
Keplerian angular frequency at $s=s_0 \equiv 1$
\begin{equation}
\label{time-unit} 
[t] = \sqrt{\frac{GM}{s_0^3}} = \Omega_0^{-1},
\end{equation}
which gives an orbital period $P=2\pi$ at $s_0$ in absence of a
global pressure gradient.

The unit of velocity \[[\vv{u}]=[x]/[t]=\Omega_0 s_0\]is therefore the local
Keplerian speed at $s_0$. The sound speed is set accordingly, through the Mach
number (see equation \ref{mach}). Density is measured relative to the initial
density of the box [$\rho$] = $\rho_0$. 

The unit of magnetic field follows from the Alfv\'en speed, \[[\vv{B}] =
\Omega_0 s_0\sqrt{\mu_0\rho_0}.\] It follows from this that the unit of
magnetic vector potential is \[[\vv{A}] = [\vv{B}][x]=\Omega_0
s_0^2\sqrt{\mu_0\rho_0}.\] 
\begin{table}
\label{physicalunits}
\caption[]{Conversion between code and physical units}
\begin{center}
\begin{tabular}{l r} \hline\hline

Quantity         & Physical Unit                               \\\hline
Length           &     5.2 AU  ($=7.8 \times 10^{11}\,{\rm m}$) \\
Density          & $\xtimes{2.0}{-8}\, {\rm kg\,m^{-3}}$       \\
Velocity         & $\xtimes{1.31}{4}$ \,${\rm m\,s}^{-1}$\\
Energy           & $\xtimes{1.60}{36}\,{\rm J}$                \\ 
Pressure, Stress & 3.41 Pa                               \\
Time             & 1.89 yr ($=6.0 \times 10^7\,{\rm s}$)   \\
Magnetic Field   & $\xtimes{2.07}{-3}$\, T          \\
Viscosity        & $\xtimes{1.02}{16}\,{\rm  m^{2}\,s^{-1}}$   \\ 
Mass             & $\xtimes{4.73}{-3}$\,M$_\odot$              \\ 
Mass accretion rate & $\xtimes{2.51}{-3}$\, M$_\odot$ yr$^{-1}$\\ 

                 &                                             \\
Domain Size\,($L_s$,$L_z$) & 2-13 AU, $\pm$1.3 AU              \\
Resolution ($\Delta\,x$)& 0.08 AU                              \\\hline
\end{tabular}
\end{center}
\end{table}

As the simulation is dimensionless, it scales with the choice of physical
units. By assuming that $s_0$ is the semi-major axis of Jupiter, $a_{\rm J} =
5.2 {\rm AU}$, and considering the typical density of the minimum mass solar
nebula at that location, $\rho_0 \approx \xtimes{2}{-8} {\rm kg\,m^{-3}}$, the
physical units corresponding to the employed code units are listed in Table~1.

\subsection{Initial conditions}

We use a Cartesian box with a spatial range $x,y\in[-2.6,2.6]$, and
$z\in[-0.26,0.26]$
(see Table 1 for a conversion of the units used to physical units). The number
of cells is usually $N_x$=$N_y$=320, $N_z$=32. This ensures that
$\Delta\,x$=$\Delta\,y$=$\Delta\,z$, i.e, all cells are cubes of the same size.
However, for some models we double the resolution in the vertical direction in
order to resolve faster growing wavelengths of the MRI.  Doubling the
resolution in $x$ and $y$ would keep the cells cubic, but the already expensive
computational costs would become unpractical without yielding any other major
advantage. We therefore keep it at 320$\times$320 and introduce anisotropic
hyperdiffusivity to treat the non-cubic cells (see
\App{ch:hyperdissipation}). 

As stated before, we use the ideal gas law approximation to evaluate the
pressure. The sound speed is set as a power law
\begin{equation}
  c_s=c_{s_0} s^{-q_{_T}/2}.
\end{equation}
We usually set $q_{_T}$=1, so that the Keplerian flow has a constant Mach
number ${\mathcal M}$

\begin{equation}
\label{mach}
{\mathcal M} =  \frac{\Omega_K s}{c_s} \equiv {\rm const.},
\end{equation}
where $\Omega_K$ is the ``cylindrical'' Keplerian angular velocity
profile
\[
\Omega_K^2 =  \frac{GM_\star}{s^3}. 
\]
The Mach number is seen to be the inverse of the aspect ratio $h=H/r$, where
$H=c_s/\Omega$ is the pressure scale height. We checked the evolution and
saturated state of the turbulence for Mach numbers of 5, 10 and 20. 

We also perform simulations with radially-varying ${\mathcal M}$ where the
sound speed follows a steeper power law, with $q_{_T}$=2.

Accretion disks exhibit a radial density gradient, but this gradient arises due
to accretion itself (Shakura \& Sunyaev 1973). Therefore, we initialize the
midplane density at a constant value $\rho_0$, in order to understand the role
of the stresses in generating the density gradient. 

We start our models in strict equilibrium between gravity, global (thermal
and magnetic) pressure gradients and centrifugal forces,
\begin{equation}
  \label{MHStatic-equilibrium}
  \Omega^2 = \Omega_{\rm K}^2 + \frac{1}{s\rho} \pderiv{}{s} \left(p + \frac{B^2}{2 \mu_0}\right).
\end{equation}

\subsection{Boundary conditions}

In this work, we compute models with and without an inner boundary to
quantify
the advantages/drawbacks of such a feature in a Cartesian grid. As for the
external boundary, the box limits at $x=\pm2.6$ and $y=\pm2.6$ do not
correspond to the physical boundaries of the problem. Indeed, the dynamically
evolving disk encompasses a cylinder inside of \rext$=2.5$. The frozen regions
outside of this cylinder play the role of ghost rings in cylindrical codes.

After evolving the dynamical equations, we set the time derivatives of all
variables to zero in the region outside \rext$=2.5$. As the variables cannot
evolve outward of \rext, being effectively frozen in this region, the ``real''
boundary conditions of the box (e.g., open, reflecting) do not matter if this
freezing boundary condition is used. 

\begin{figure}
\label{borderfigure}
\begin{center}
  \resizebox{8cm}{!}{\includegraphics{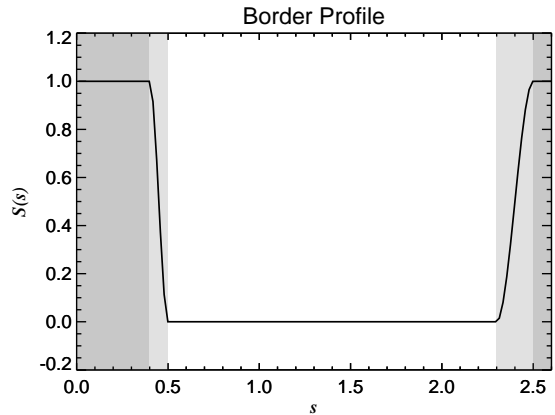}}
\end{center}
\caption[]{Border profile of the simulations. The step function 
(Eq.~\ref{border-profile}) is applied to the derivatives of the
dynamical variables. The gas is free to evolve between $s$=0.5 and $s$=2.3, is
slowed down between [0.4,0.5] and [2.3,2.5] (light-shaded areas) and is
effectively frozen at $s<0.4$ and $s>2.5$ (dark-shaded areas).}
\end{figure}

To avoid numerical instabilities due to this abrupt jump from frozen to
evolving regions, we apply a buffer zone to the derivatives of the variables,
that smoothly drives the variable $X$ to a desired value $X_0$ in a timescale
$\tau$, such that
cp \begin{equation}
\label{border-profile}  
  \frac{\partial{X}}{\partial{t}}=-\frac{X-X_0}{\tau} \mathcal{S}(s).
\end{equation}where $\mathcal{S}(s)$ is the uniquely defined fifth-order step function 
that is 1 at the domain boundary and 0 at the interior boundary of the 
buffer zone while maintaining continuous second order derivatives 
(Dobler, private communication). Its shape is visualized in Fig.~1.
We usually take the driving term $X_0$ to be the initial condition of the variable, and the driving
time $\tau$ being the Keplerian period $2\pi/\Omega_{\rm K}$, the dynamical timescale
of the disk. The effect of this border profile is to smooth the transition
between the evolving disk and the frozen regions of the grid, thus preventing
large gradients and discontinuities that would otherwise arise. As the gas flow
is symmetric in the vertical direction, the vertical boundary condition is set
to periodic for the purpose of simplicity.

For the runs without an inner boundary, we smooth the quantities containing
singularities by replacing 
\begin{equation}
  \label{smoothing}
  s^{-n} \Rightarrow (s^2 + b^2)^{-n/2}.
\end{equation}
In practice, it is applied to the angular frequency $\Omega$, the gravitational
potential $\Phi$ and the sound speed $c_s$. We usually take $b=0.1$, so the
smoothed gravitational potential deviates from the Newtonian by less than 5\%
at $s_{\rm int}=0.4$. The physical domain thus runs from \rint to \rext.

For runs with an inner boundary, we apply inside \rint the same freezing as
used outside \rext. The $N$-body particle code does not participate in the
freezing, so although the star lies in a region of frozen gas, it is allowed to
move. 

As the gas is frozen in the inner and outer parts, the information about the
flow in this region is not of interest. Therefore, we exclude these regions
from the time-step calculation. As their time derivatives are set to zero at
the end of the time-step, they cannot violate causality.

In principle, we could set \rint as close to zero as possible (by not using
smoothing but retaining an inner boundary), in order to study the processes
that happen in the immediate vicinity of the star, like winds, the magnetic
cavity and surface accretion (von Rekowski \& Piskunov 2006). However, due to
the increasing Keplerian velocity in the advection and the decreasing
resolution of the orbits, non-axisymmetric wave modes (particularly the $m$=4 mode) build up
in the inner disk as we try to push \rint $\rightarrow 0$. The density fluctuations resulting from the excitation of
these modes lead to numerical instabilities.

Finally, the magnetic potential follows
the same boundaries as described above. This would be a problem if we solved
for the actual magnetic field, as sinks or sources of magnetic flux
imply the presence of open
magnetic loops (monopoles). By solving for the magnetic vector potential we do not face such problems. 

The solid particles obey different boundary conditions, explained in Sect.~5.

\section{Influence of free parameters}

In order to clarify the influence of the numerical scheme and the
approximations made, a series of non-magnetic 2D models were computed, with and
without planets. The grid being Cartesian, all our simulations span the whole
azimuthal domain. We usually evolve the simulations up to 100 orbits at $s_0$,
which corresponds to $\approx$25 orbits at the outer edge of the disk and
$\approx$400 orbits at its inner edge.  

\subsection{Viscosity}

Explicit hyperviscosity and hyperdiffusion induce dissipation primarily near
the grid scale, replacing the usual 2$^{\rm nd}$ order Laplacian terms. A
visual picture of the difference between using the two types of viscosity is
seen in Fig.~2. The first model was computed with a Laplacian viscosity
$\nu_1=\ttimes{-3}$. The radial inflow is significant, and as the outer frozen
region behaves like an infinite reservoir of matter, the total mass inside the
disk keeps on rising as matter flows in from this reservoir.  The radial
density profile soon starts to deviate from the flat initial condition .  Shown
in the figure is the density profile after 100 orbits at $s_0$ with
$\nu_1=\ttimes{-3}$. When using hyperviscosity of similar strength at the grid
scale, i.e., $\nu_3=\ttimes{-10}$, the overall flow shows no significant
deviations from the initial conditions. 

The simulation shown using Laplacian viscosity was computed without an inner
boundary, using a softened stellar potential with $b$=0.1. Using the damping
and freezing profile described in Sect. 2.7, the density is not allowed to
deviate much from the initial condition at the boundaries. In the physically
evolving part of the disk, however, a density profile of exponent $s^{-0.4}$
evolves.

\subsection{Mass diffusion}

To evaluate the influence of mass diffusion, we simulate a laminar disk
with
constant Laplacian viscosity $\nu_1=\ttimes{-5}$ in the presence of a
gap-opening Jupiter-mass planet. We performed runs of resolution $320 \times
320$ with hyperdiffusion coefficients ranging from $D_3=\xtimes{5}{-11}$ to
$\ttimes{-14}$. After a hundred orbits, time enough for the planet to open a
deep gap, the density profiles are plotted in Fig.~2a, where a run with
resolution $640 \times 640$ without explicit diffusion is shown for comparison. 

It is seen that $D_3=\xtimes{5}{-11}$ constitutes too much diffusion, as the
gap is significantly altered. As less diffusion is used, the shape of the gap
monotonically approaches the one recovered in the higher resolution run. The
walls of the gap are fairly well reproduced for lower diffusion, but its bottom
is always shallower even for the lowest coefficient used ($D_3=\ttimes{-14}$). 

Judging from the gap alone, one could in principle use no diffusion at all, but
the inner disk suffers depletion for low diffusion regimes, due to the
non-conservative nature of the numerical scheme. Even the higher resolution run
seems to have lost mass due to the lack of explicit diffusion.

We adopt a hyperdiffusion coefficient of $D_3=\xtimes{5}{-12}$ as the best
compromise between the need for preserving material in the inner disk and for
reproducing the overall shape of the gap. Requiring Schmidt and magnetic
Prandtl numbers of 1 at the grid scale, we set hyperviscosity and
hyperresistivity to the same value.

\begin{figure}
\label{viscevolution}
\begin{center}
  \resizebox{8cm}{!}{\includegraphics{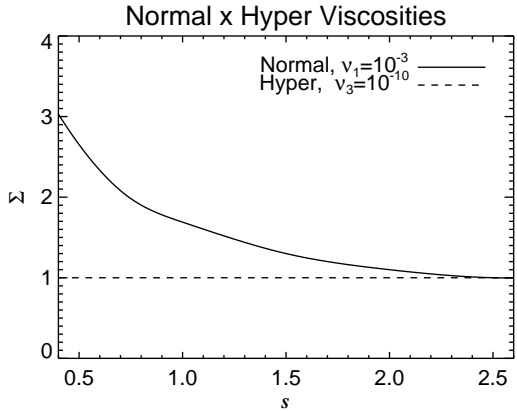}}
\end{center}
\caption[]{The radial density profile after 100 orbits for Laplacian viscosity
$\nu_1=10^{-3}$ is compared to the profile obtained by using sixth-order
hyperviscosity ($\nu_3=\ttimes{-10}$) of same strength in the small scales. In
the hyperviscous case, the global flow is unaffected.  Note that the frozen
regions behave like infinite reservoirs of matter. The power law describing the
resulting density profile for normal viscosity is very close to the $s^{-0.5}$,
as expected for constant viscosity (see Pringle 1981 and references therein)}
\end{figure}

\begin{figure}
\label{diffusive-gap}
\begin{center}
  \resizebox{7.5cm}{!}{\includegraphics{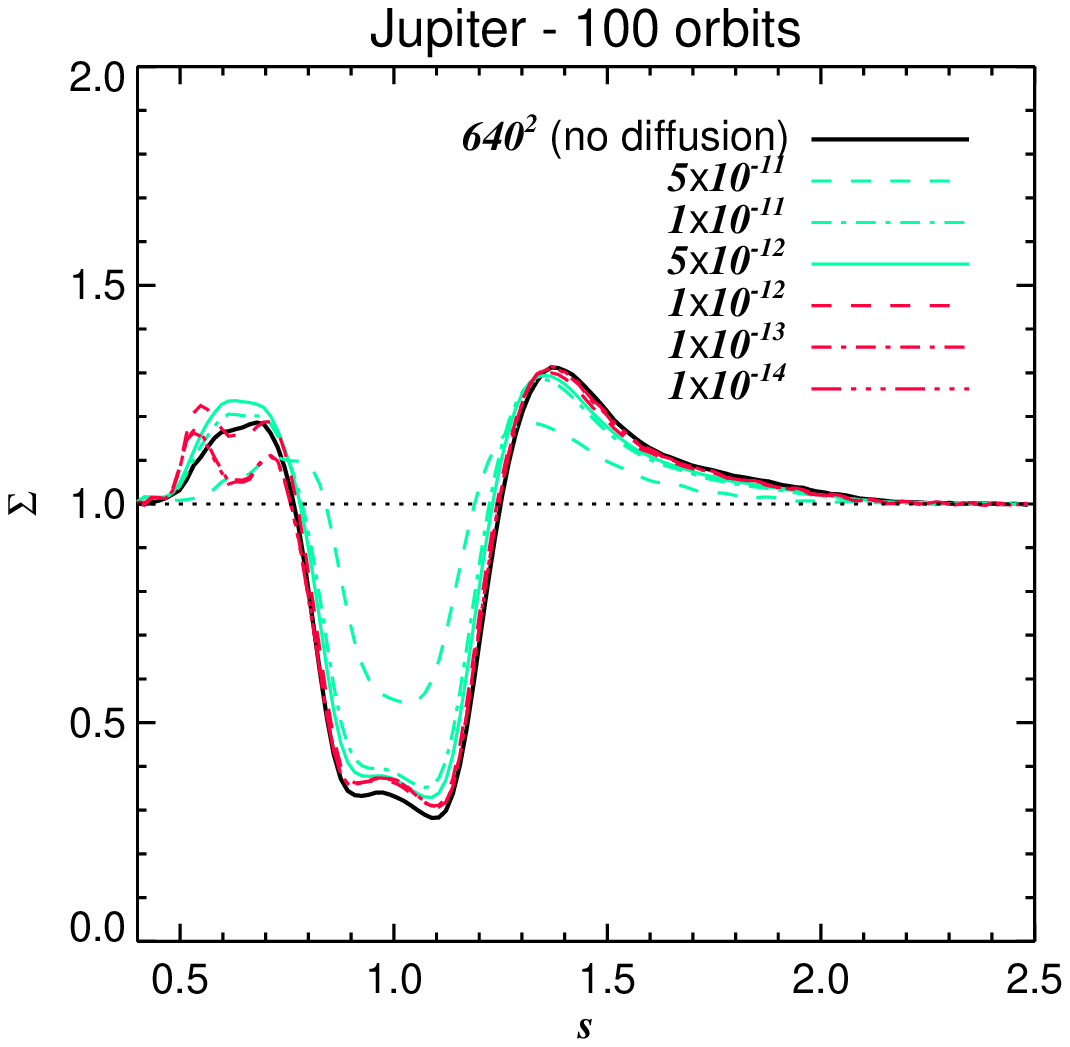}}
  \resizebox{7.5cm}{!}{\includegraphics{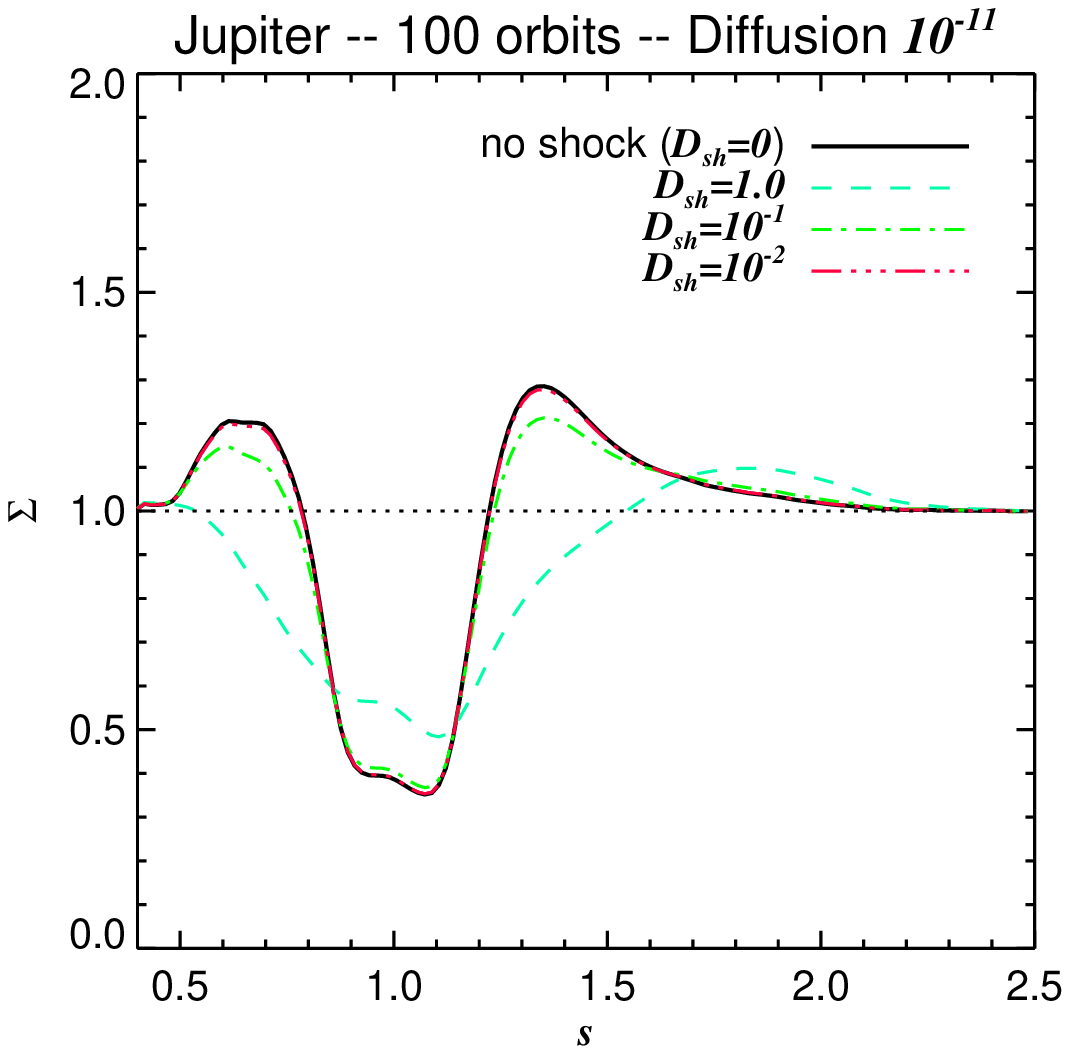}}
\end{center}
\caption[]{{\it Upper Panel}. The gap carved by a 1 $M_J$ planet in a 2D disk
reveals the influence of explicit diffusion in the calculations. The inner disk
loses mass depending on the amount of diffusion. The value of
$D_3=\xtimes{5}{-12}$ seems to ensure mass conservation in the inner disk yet
not distorting the shape of the gap. The solid line represents a $640\times640$
run without diffusion, for comparison. Resolution is $320\times320$ otherwise.
\\ {\it Lower Panel}. Same but with hyperdiffusion set to $D_3=\ttimes{-11}$
and varying the shock diffusion coefficient from $\ttimes{-2}$ to $1$.}
\end{figure}

\subsection{Shocks}

Shock viscosity and shock diffusion are needed for two reasons: ({\it a.}) to
stabilize the flow near the shock-generating particles in runs with planets and
({\it b.}) to treat eventual supersonic motion in the turbulence (arising when
the disk is exposed to a strong net vertical field), in which case shock
resistivity is also included.

In Fig.~2b we show gap-opening runs with fixed hyperdiffusion coefficient $D_3$
but varying the shock diffusion coefficient. From the continuity equation, one
can tell that the effect of shock diffusion is to slow down the time evolution
of density by smearing out any large divergences. Indeed, one sees that after a
hundred orbits, a shock diffusion coefficient of $1$ fails to reproduce the
shape of the gap as compared to the higher resolution run without shock
diffusion, while $\ttimes{-1}$ shows less accumulation than expected in the
Lindblad resonances, also seen as compared to the higher resolution run. We
therefore use shock diffusion of $\ttimes{-2}$ for the turbulent runs.  The
flow around a high-mass planet, however, could only be stabilized with a shock
viscosity of 1.

Shock resistivity is also used when the run involves the magnetic potential.
The value used was not tuned in 2D runs like shock diffusion and shock
viscosity. Instead, in the presence of turbulence, we simply checked what was
the lowest shock resistivity coefficient that did not lead to numerical
instabilities for model A (see Table~2), finding that it is of the order of
unity, like the shock viscosity.

\subsection{Non-turbulent runs}

To verify the numerical stability of the model in the absence of physical
turbulence, we perform tests for cases where the turbulence is not supposed to
be present. In these runs, we monitor the evolution of the mass inflow rate
$\dot{M}$ defined as the 1D radially dependent surface integral over a
surface $\mathcal A$ which is a cylinder at a radial distance $s$ from the
origin
\begin{eqnarray}
\dot{M}(s)&=&\varoiint_{\mathcal A} \rho\,\vv{u}\cdot\hat{\vv{n}}\,\de{A} \\
&=& 2\pi s \int_{_{-L_z/2}}^{^{\,\,L_z/2}} \rho(s,z) u_s(s,z)\,\de{z}.
\end{eqnarray}
We see that in a 2D laminar model, the mass inflow rate is constant 
through the radial domain once a steady flow is achieved (which simply 
states that mass is conserved). We thus define the mass accretion
rate as the mass inflow rate across the inner boundary, 

$\dot{m}=\dot{M}\vert_{_{s=0.4}}$,

meaning that after crossing this boundary, the matter is considered lost
(accreted).

We also measure the kinetic alpha parameter of turbulent viscosity,
defined as\[\alpha_R = \frac{2}{3}\frac{R^{s\phi}}{\rho c_s^2},\]where
$R^{s\phi}=\overline{\rho\delta u_s \delta u_\phi}$ is the Reynolds stress; and
its magnetic counterpart\[\alpha_M = -\frac{2}{3}\frac{M^{s\phi}}{\rho
c_s^2},\]where $M^{s\phi}= \mu_0^{-1}\overline{\delta B_s \delta B_\phi}$ is
the Maxwell stress.

\begin{figure}
\begin{center}
\resizebox{8cm}{!}{\includegraphics{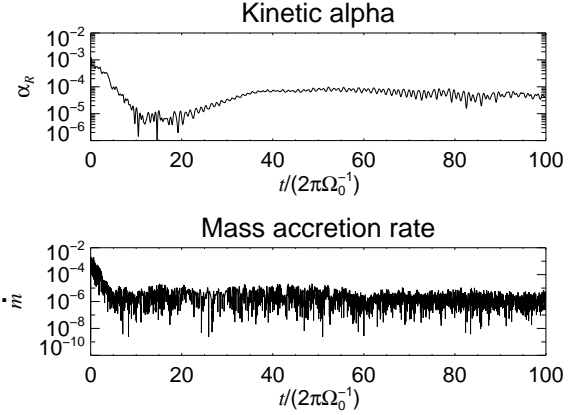}}
\end{center}
\caption[]{Without solving the dynamical equations for the magnetic field, even
initially vigorous random motions of $u_{\rm rms}=0.2 c_{s_0}$ die out rather
quickly. The plots show the time evolution of the globally averaged Reynolds
stress and mass accretion rate. Time is quoted in orbits at $s_0$.}
\label{turbdie}
\end{figure}

In Fig.~\ref{turbdie} we show a 2D run where the velocities were perturbed with
noise of $u_{\rm rms} = \ttimes{-2}$, but the induction equation is not
solved. The turbulence dies out so fast that even before ten orbits at $s_0$
the flow is already smooth, showing that the numerical scheme does not
spuriously generate or sustain turbulence.

A 3D cylindrical run in which we add a vertical net field of strength
$B_0=\ttimes{-3}$ (dimensionless), corresponding to plasma $\beta$=5000 at
$s_0$, 12500 at \rint and 2000 at \rext (see Eq.~[\ref{plasma-beta}]), but
where the initial flow is not perturbed by noise, does not develop turbulence.
We also tested if the MRI would develop in a disk seeded only with noise in
both the velocity and magnetic potential ($u_{\rm rms} = A_{\rm rms} =
\ttimes{-4}$). There is a short growth in magnetic energy, presumably due to
reconnection of the field lines, but without a structured field to maintain the
turbulence the Reynolds and Maxwell stresses quickly level down to zero
as viscosity and resistivity smooth the imposed noise.  

\begin{table*}
\label{diskmodels}
\caption[]{Cylindrical turbulent disk models.} 
\begin{center}
\begin{tabular}{lcc cc ccc ccc ccc ccc} \hline \hline
&\multicolumn{8}{c} {\sc Parameter} & & \multicolumn{7}{c}{\sc Results}\\\cline{2-9}\cline{11-17}
\multirow{2}{*}{\sc Run}&\rint&$B_0$&$c_{s_0}$&$q_{_T}$&$\beta_0$&$N_z$&$\chi_{\rm sh}$&$N_p$&&$R^{s\phi}$&$-M^{s\phi}$&$\alpha_R$&$\alpha_M$&$B_{\rm rms}$&$\beta_t$&$\delta_t$\\
&&\tiny{($\xtimes{}{3}$)}&&&&&&&&\tiny{($\xtimes{}{5}$)}&\tiny{($\xtimes{}{5}$)}&\tiny{($\xtimes{}{3}$)}&\tiny{($\xtimes{}{3}$)}&\tiny{($\xtimes{}{3}$)}& &\tiny{($\xtimes{}{3}$)}\\\hline

\multicolumn{17}{c}{\multirow{2}{*}{Uniform Field $B_z$ }}\\\\\hline
A &0.4&1&0.05&1&5000&32&1&$\ttimes{6}$&& 
\mm{0.24}{0.04}&\mm{1.5}{0.3}&\mm{0.9}{0.2}&\mm{6}{1}&\mm{17}{9}&\mm{13}{3}&\mm{7}{1}\\
B &0.4&1&0.10&1&20000&32&1&...&&
\mm{1.0}{0.2}&\mm{2.5}{0.3}&\mm{0.7}{0.1}&\mm{1.8}{0.2}&\mm{16}{5}&\mm{65}{7}&...\\
C &0.4&1&0.20&1&80000&32&1&...&&
\mm{4}{1}&\mm{5.3}{0.8}&\mm{0.9}{0.2}&\mm{1.3}{0.2}&\mm{26}{7}&\mm{81}{7}&...\\
A\,2&0.0&1&0.05&1&5000&32&1&...&&
\mm{0.20}{0.04}&\mm{1.3}{0.1}&\mm{0.7}{0.1}&\mm{4.7}{0.3}&\mm{17}{5}&\mm{13}{1}&...\\
B\,2&0.0&1&0.10&1&20000&32&1&...&&
\mm{0.9}{0.2}&\mm{2.6}{0.3}&\mm{0.8}{0.1}&\mm{2.1}{0.2}&\mm{20}{11}&\mm{43}{13}&...\\
C\,2&0.0&1&0.20&1&80000&32&1&...&&
\mm{5}{3}&\mm{5}{1}&\mm{1.2}{0.8}&\mm{1.2}{0.3}&\mm{22}{9}&\mm{116}{19}&- \\\hline

 \multicolumn{17}{c}{\multirow{2}{*}{Radially Varying Field $B_z$}}\\\\\hline
D&0.4&20&0.10&2&12&64&2&$\ttimes{6}$&&
\mm{22}{2}&\mm{78}{7}&\mm{25}{1}&\mm{87}{3}&\mm{71}{9}&\mm{4}{1}&\mm{140}{10}\\
E&0.4&20&0.20&2&50&64&2&...&&
\mm{35}{7}&\mm{87}{15}&\mm{13}{3}&\mm{30}{5}&\mm{60}{23}&\mm{11}{1}&...\\

Dw&0.4&5&0.10&2&750&64&2&...&&
\mm{4.1}{0.4}&\mm{13}{2}&\mm{5.4}{0.6}&\mm{17}{2}&\mm{28}{10}&\mm{13}{2}&...\\

Ew&0.4&5&0.20&2&3000&64&2&...&&
\mm{13}{4}&\mm{27}{2}&\mm{4}{1}&\mm{8.7}{0.7}&\mm{36}{11}&\mm{33}{3}&...\\\hline

 \multicolumn{17}{c}{\multirow{2}{*}{Uniform Field $B_\phi$}}\\\\\hline

F&0.0&30&0.05&1&5.5&32&2&...&&
\mm{0.7}{0.1}&\mm{2.9}{0.4}&\mm{2.5}{0.4}&\mm{11}{1}&\mm{12}{4} &\mm{24}{2}&...\\
G&0.0&30&0.20&1&90&32&2&...&&
\mm{5}{2}&\mm{18}{6}&\mm{0.9}{0.4}&\mm{3.3}{1.4}&\mm{28}{11} &\mm{107}{22}&...\\\hline
\hline
\end{tabular}
\end{center}
\end{table*}

\section{Cylindrical disk runs}

For the main simulations in this paper we consider flat vertical profiles for
the gravity field. Such an approximation is called a ``cylindrical'' disk and
has been often used in order to study the MRI (e.g.,
Armitage 1998, Hawley 2001). The vertical gravity $g_z = -\Omega^2 \vv{z}$ is 
switched off so that, physically, the star is no longer a point mass at 
$r$=0, but a rod at $s$=0 extending through the length of the $z$-axis. 
In such a setup, the pressure scale height $H$ has no hydrostatic meaning, 
being only a way to write the temperature profile of the disk. We performed 
simulations with non-zero net flux magnetic fields $\vv{B}=B_0\hatz$ and 
$\vv{B}=B_0\hatphi$. Models with a radially varying vertical field 
proportional to $\Omega(s)$ were also computed.

For the turbulence to develop, the unstable wavelengths of the MRI must be
resolved. The characteristic vertical wavelength $\lambda_{\rm BH}$ of
the hydromagnetic turbulence is given by (Balbus \& Hawley 1991, 1998)
\begin{equation}
\label{balbus-hawley}
  \lambda_{\rm BH} = 2\pi \frac{\va}{\Omega},
\end{equation}
where $\va$ is the Alfv{\'e}n speed 
\begin{equation} 
\label{alfven}
  \va = \frac{B}{\sqrt{\mu_0\rho}}. 
\end{equation} 
The turbulence will be present as long as the critical wavelength $\lambda_c$
is resolved. This wavelength is $\lambda_c=\lambda_{\rm BH}/\sqrt{3}$, whilst
the most unstable wavelength of the MRI is $4\lambda_{\rm BH}/\sqrt{15}$ 
(Balbus \& Hawley, 1991). The plasma $\beta$ parameter - the ratio of thermal to
magnetic pressure - can be expressed in terms of the sound and Alfv\'en speed,
by writing the magnetic pressure $P_M=B^2/(2\mu_0)$ in terms of the Alfv\'en
speed $P_M = B^2/(2\mu_0) = \rho \va^2/2$, giving
\begin{equation}
  \label{plasma-beta}
  \beta = \frac{2c_s^2}{\va^2}.
\end{equation}
The constant $B_0$ is usually set to $\ttimes{-3}$, but runs varying the field
from $\ttimes{-4}$ to $\ttimes{-1}$ were also studied. Although the runs
reported here are in the local isothermal approximation, we also varied the
initial sound speed at $s_0$, $c_{s_0}$, in order to check how the resulting
turbulent viscosity depends on the global gas pressure.

The parameters of the cylindrical models presented here are specified in
Table~2.

\subsection{Constant vertical field - model A}

\begin{figure*}
\begin{center}
\resizebox{13.75cm}{!}{\includegraphics[angle=90]{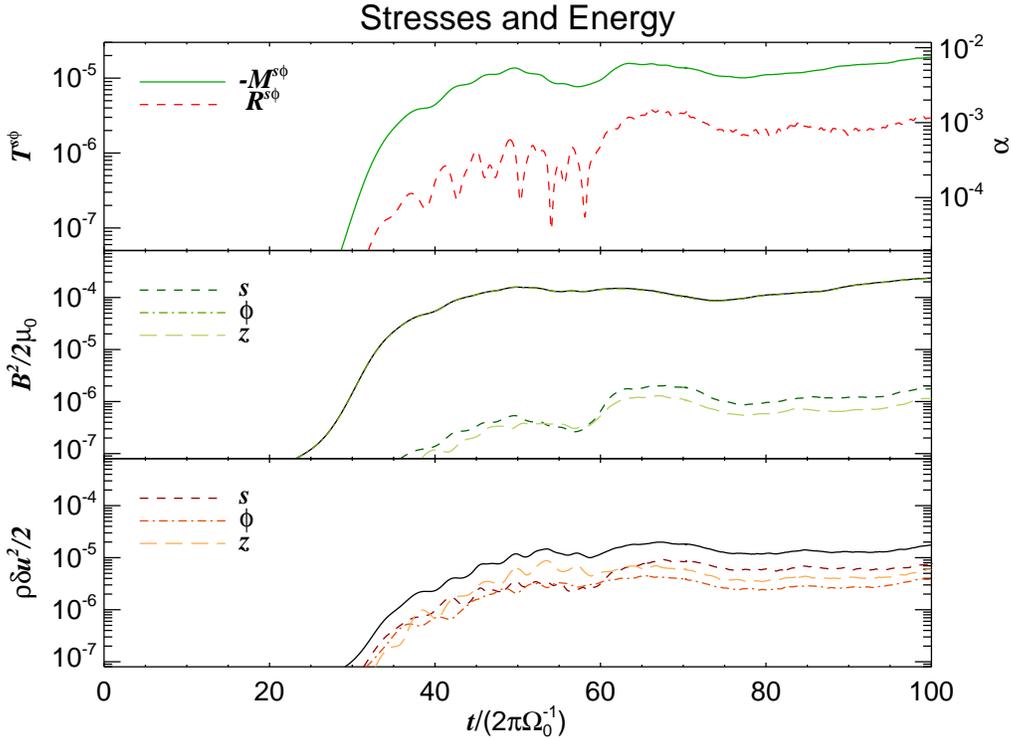}}
\end{center}
\caption[]{Time evolution of the turbulence for model A (constant net flux
vertical field $\vv{B}=\ttimes{-3} \hatz$ and constant Mach number).
The top panel shows the evolution of the $s\phi$-component of
the Maxwell and Reynolds stresses, while the middle and bottom panels show the
evolution of magnetic and kinetic energy, respectively. The units are given in
Table~1. The solid lines in the two bottom panels show the total energy. The
toroidal field dominates the magnetic energy to the point that the energy in
the azimuthal component can barely be distinguished from the total energy. The
kinetic energy is more evenly distributed among the three dimensions, but the
turbulence is not isotropic. The radial component shows 1.5 times more energy
than the vertical and 2 times more than the azimuthal. Time is quoted in orbits
at $s_0$.}
\label{fiducialcyl}
\end{figure*}

The evolution of the turbulence in a fiducial run with a net vertical flux of
strength $B_0=\ttimes{-3}$ and temperature profile corresponding to
$c_{s_0}=0.05$ is shown in Fig.~\ref{fiducialcyl}. The absolute value of the
Maxwell stress at saturation is always larger than the Reynolds stress, but the
latter fluctuates more strongly.

The minimum ratio of stresses is
$-M^{s\phi}/R^{s\phi}$=3 (at t$\approx$75 orbits), but it reaches as much as 100 (at t$\approx$58
orbits). After 75 orbits, the average ratio of Maxwell to Reynolds stress is
around 5. 

\begin{figure}
\begin{center}
\resizebox{8cm}{!}{\includegraphics{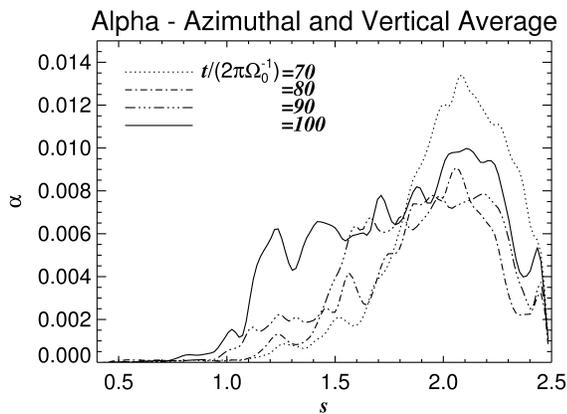}}
\end{center}
\caption[]{Radial structure of the total alpha parameter $\alpha_R+ \alpha_M$
for model A . The turbulence starts from the outer disk, where plasma $\beta$
is smaller. The different curves correspond to snapshots at 70 (dotted), 80
(dot-dashed), 90 (dot-dot-dot-dashed), and 100 orbits (solid), after saturation
is reached. The variability is not monotonic, but highly fluctuating.}
\label{radalphacs5}
\end{figure}

In agreement with previous shearing box simulations (Brandenburg et al.  1995,
Hawley et al. 1995, Johansen \& Klahr 2005), global disks (Hawley 2001, Nelson
2005) and analytical calculations (Balbus \& Hawley 1991), a large scale
toroidal field is seen to form, which dominates the magnetic energy, being 2
orders of magnitude stronger than the radial and vertical fields. Indeed, one
can barely distinguish between the energy stored in the azimuthal field and the
total magnetic energy. The kinetic energy is more evenly distributed, but it is
not isotropic. The radial component accounts for 45\% of the total
energy, being $\approx$1.5 bigger than the vertical and twice as big as the
azimuthal. The radial structure of the alpha parameter is seen in
Fig.~\ref{radalphacs5}.  The outer disk is more turbulent due to the smaller
values of plasma $\beta$ when compared to the inner disk. 

Following the time evolution of the turbulence in the midplane, it is seen that
different regions of the disk reach saturation at different times. The
turbulence starts from the outer disk, propagating inwards. It is expected since, as $\Omega$ decreases with radius, a uniform 
field implies a Balbus-Hawley wavelength that increases with distance from 
the star. As longer wavelengths - comparable to the length of the box - 
are easily resolved, the outer disk goes turbulent first. It is seen that 
inside $s_0$=$1$ the disk did not go
turbulent. In this model, the Mach number is constant, ${\mathcal M}$=$20$,
with a constant field, $B$=$\ttimes{-3}$ through the whole domain,
corresponding to plasma $\beta$ running from  2000 at \rext and 12500 at \rint.
The magnetic field determines the value of the critical wavelength, which
ranges from $\lambda_c$=0.002 at \rint to $\lambda_c$=0.025 at \rext.  As we
have 32 grid points in the vertical direction, the smallest wavelength 
resolved with significant accuracy (8 points) by our high-order 
finite-difference method is $L_z$/4$\approx$0.12. From the dispersion 
relation of the MRI (Hawley \& Balbus 1991), this unstable wavelength has 
a growth rate of $\approx 0.1 \Omega$, much lower than the fastest 
growing wavelength with $\omega=(3/4)\Omega$.

We also computed models where the whole disk goes turbulent (models D \& E),
but we will use these weak field disk models (models A to C2, see Table 2) 
in the next subsection for studying the behavior of the turbulence with 
thermal pressure. Although dissipative and slowly growing, the weak field 
used in these disks has one major advantage: the turbulence grows slowly 
and has less spatial variability. Therefore, the damping at the frozen 
boundaries is more gentle than in other, rapidly growing, violently 
fluctuating, disks. These issues discussed above reflect the compromise 
between keeping the disk cold while still retaining the field subthermal 
and with sufficient resolution to resolve the rapidly growing unstable 
wavelengths.

The runs with a vertical net field of $B_0=\ttimes{-4}$ or lower did not
develop turbulence as the field is weaker than needed in the presence of the
chosen dissipation parameters (Hawley \& Balbus 1991). 

\subsection{Dependence on sound speed - models B and C}

We investigate the dependence of the saturated state on the imposed sound speed
profile. We test three different sound speeds of, $c_{s_0}$=0.05, 0.10 and
0.20, corresponding to runs A, B and C. 

\begin{figure*}
\begin{center}
\resizebox{15.25cm}{!}{\includegraphics{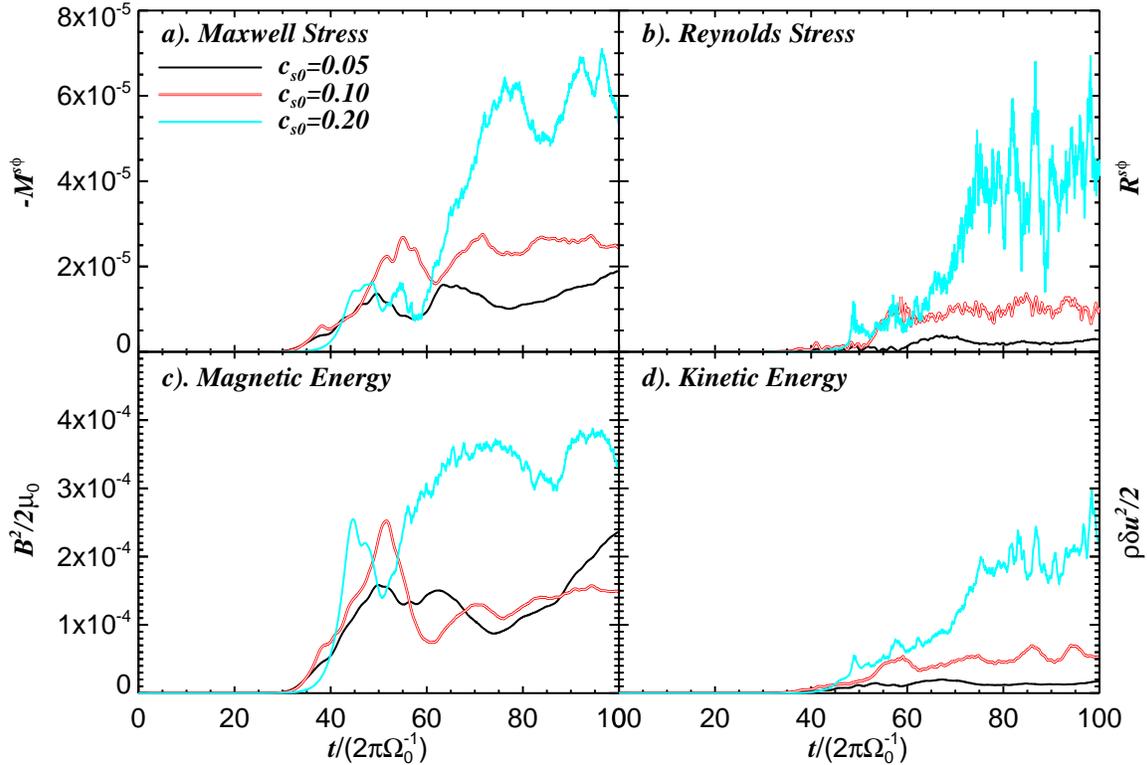}}
\end{center}
\caption[]{Time evolution of the turbulence for different sound speed profiles.
The strength of the angular momentum transport differs with sound speed. The
increase in stress observed when the sound speed is raised to $c_{s_0}=$0.20 is
dramatic. Time is quoted in orbits at $s_0$.}
\label{sound-speed}
\end{figure*}

As the runs are locally isothermal, we are mainly investigating how the
MRI responds to different radial pressure gradients.
The cold model (A) shows a weaker turbulence at saturation than the hotter
ones, as shown by the strength of the stresses (Fig.~\ref{sound-speed}a,b) and
magnetic/kinetic energies (Fig.~\ref{sound-speed}c,d). The Maxwell stress is
three times bigger for model C than for A, the colder version. Such behavior
was reported by Sano et al. (2004), indicating a power law of exponent 0.25 for
the growth of the Maxwell stress with gas pressure. Our global disk
calculations agree well with this value.

\begin{figure*}
\begin{center}
\resizebox{15.25cm}{!}{\includegraphics{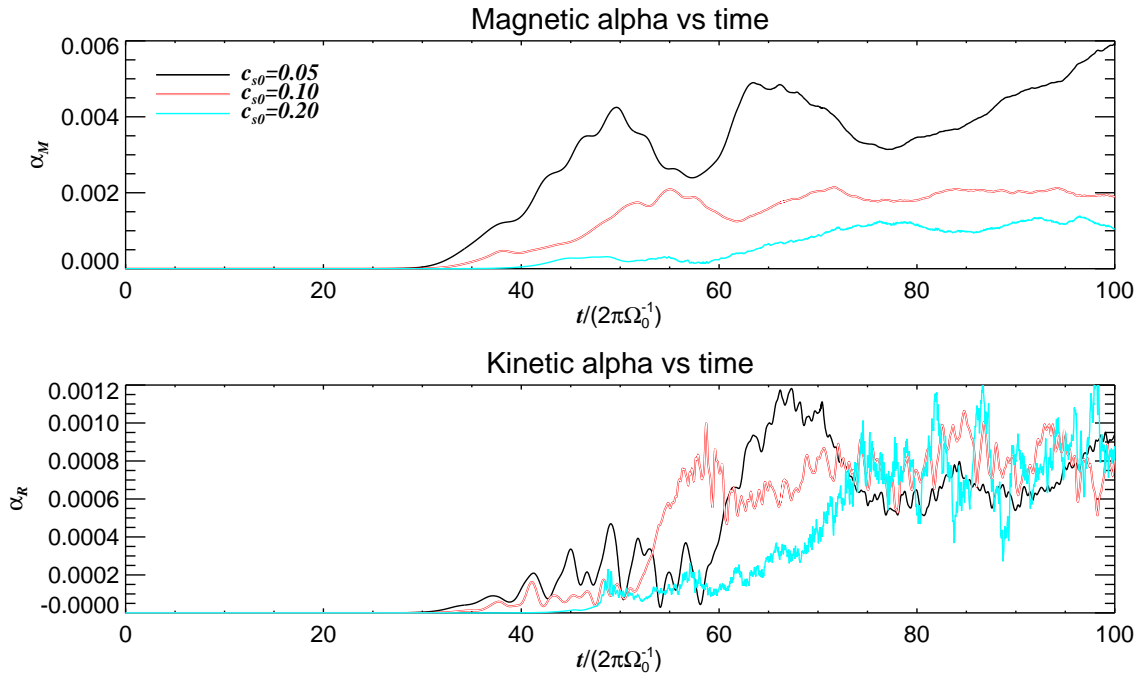}}
\end{center}
\caption[]{Evolution of the alpha parameters for different sound speed
profiles. This quantity, that measures the strength of viscosity through the
parametrization $\nu_{\rm t}=\alpha c_{\rm s} H$, decreases with sound speed
for the magnetic stresses but stays constant for hydrodynamic stresses. As seen
before, the stresses actually {\it increase} with increasing temperature, so
this decrease of the alpha parameter is due to the stresses increasing
less rapidly than the pressure.
Time is quoted in orbits at $s_0$.}
\label{alpha-cs}
\end{figure*}

\begin{figure*}
\begin{center}
\resizebox{ 8cm}{!}{\includegraphics{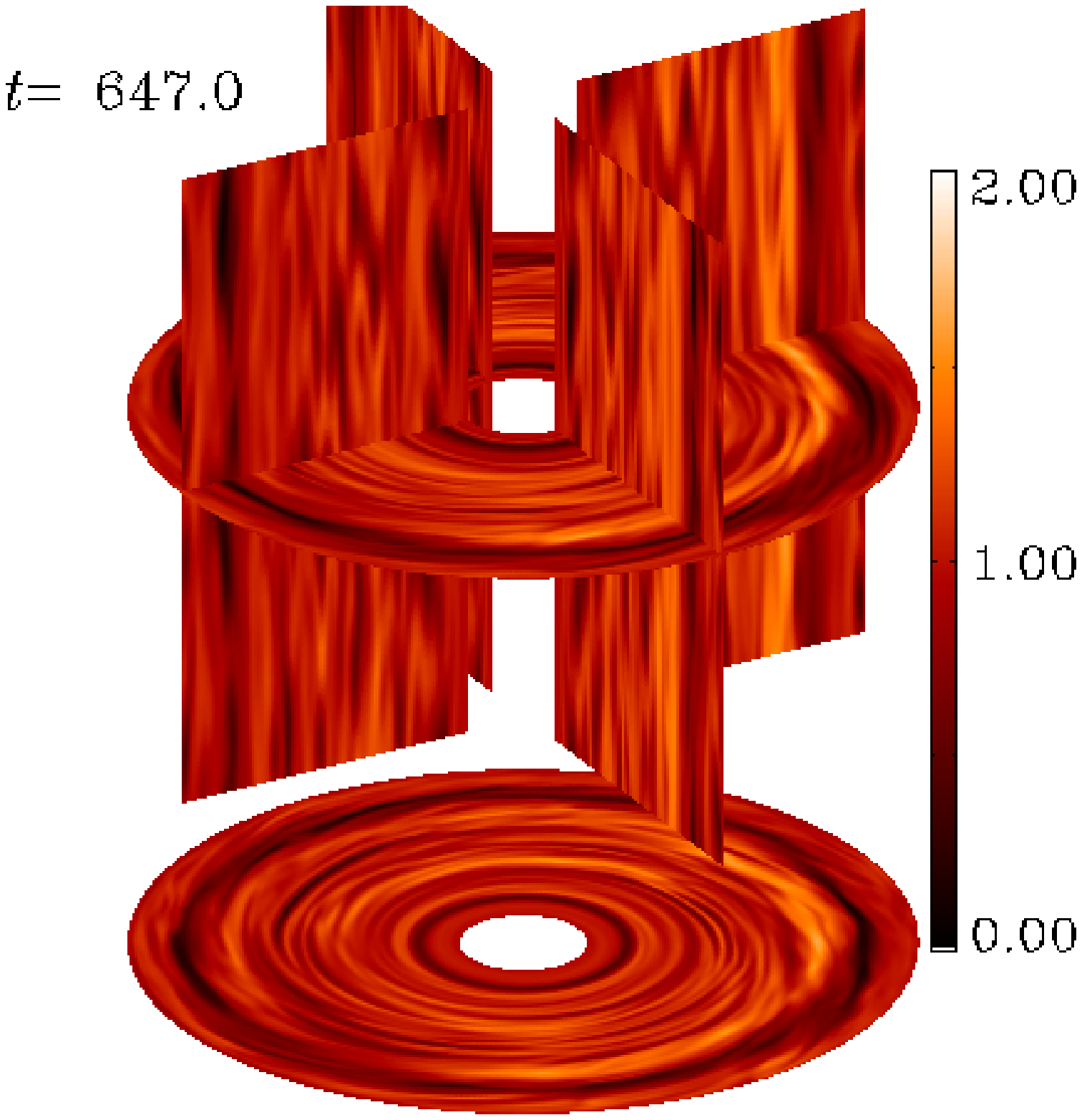}}
\resizebox{ 8cm}{!}{\includegraphics{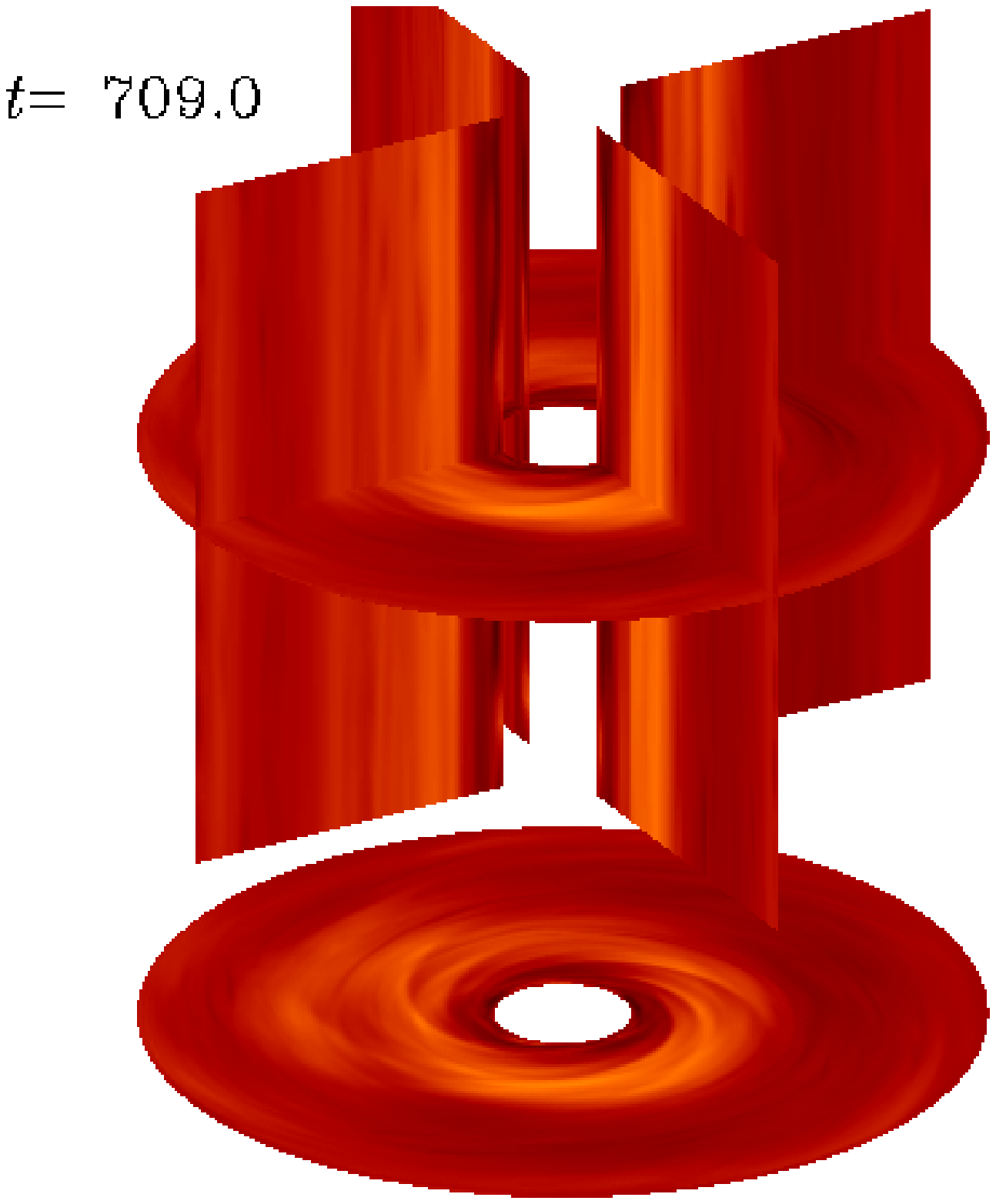}}
\end{center}
\caption[]{Density contours at selected planes $x$=$0$, $y$=$0$, and $z$=$0$ on
saturated turbulent state for sound speed profiles of $c_{s_0}=0.05$ (model A,
left panel) and $c_{s_0}=0.20$ (model C, right panel). The color code is 
the same for both figures. The stronger stresses for the hotter case lead to 
a much more effective turbulent viscosity, as seen from the steep density 
profile resulting from accretion. The vertical planes are stretched to show 
more detail than the correct aspect ratio would allow. Movies of these 
simulations can be found at {\tt http://www.astro.uu.se/$\sim$wlyra/planet.html}.}
\label{gyroplot}
\end{figure*}

\begin{figure*}
\begin{center}
\resizebox{14.75cm}{!}{\includegraphics{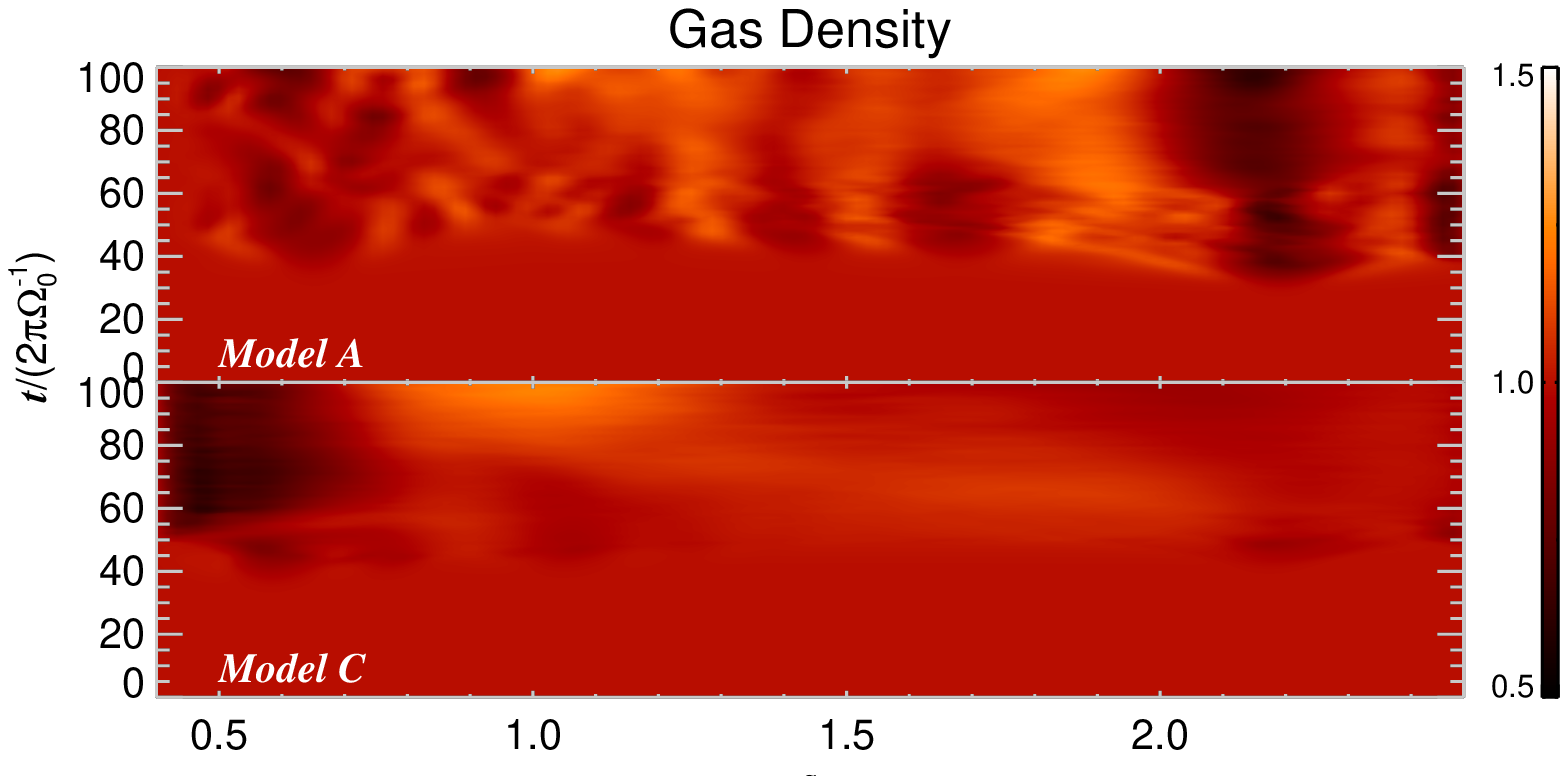}}
\end{center}
\vspace{1mm}
\begin{center}
\resizebox{13.25cm}{!}{\includegraphics{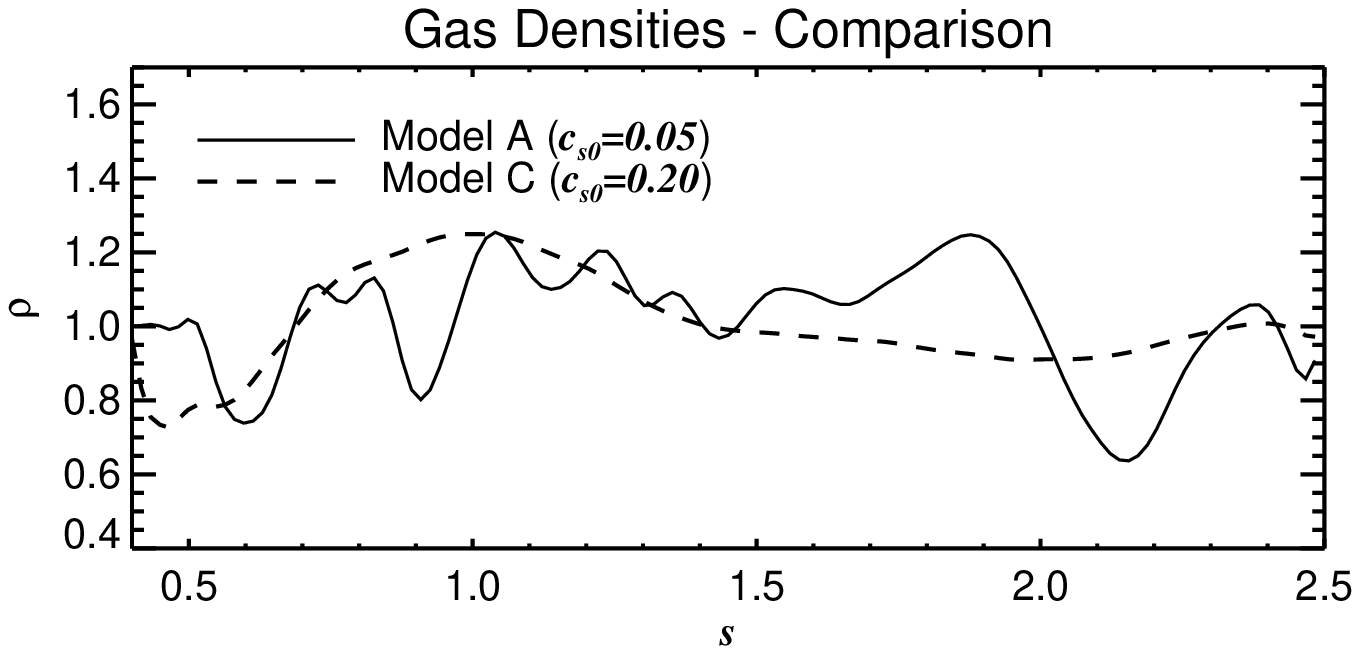}}
\end{center}
\caption[]{Radial density profiles for models A and C. The lower panels show 
 space-time plots of the vertically and azimuthally averaged density. Model A shows
  what seems to be a variability around a constant value, model C has developed 
a smoother radial density gradient. The lower panel shows the two density 
profiles at the end of the simulation.}
\label{space-gyro}
\end{figure*}

The dimensionless magnetic $\alpha_M$ parameter, which is a measure of
turbulent viscosity, decreases drastically with sound speed
(Fig.~\ref{alpha-cs}). As seen before, the stresses actually {\it increase}
with increasing temperature, so this decrease of $\alpha_M$ is due to the
stresses increasing less rapidly than the temperature. Even though alpha decreases, the effective
viscosity $\nu_{\rm t}$=$\alpha c_s H$ increases. As seen in
Fig.~\ref{gyroplot} and Fig.~\ref{space-gyro}, a centrally concentrated density profile has developed
from the initially flat configuration, a signature of mass accretion due to
turbulent angular momentum transport, as also confirmed by the measured
stresses.  The resulting density profile in model C is smoother overall, 
but the overdensities seem to be similar in average.

It is clear that alpha {\it per se} is not a good measure of viscosity.
Since $\nu_{\rm t}=\alpha c_s^2 \Omega^{-1}$, with no reference to the
Alfv\'en speed, the resulting alpha value of turbulent disks where 
the origin of the turbulence is magnetic may change with sound speed.
As most of the analyses of turbulent thin accretion
disks have focused on locally isothermal simulations using $c_s \approx 0.05$,
 such dependence of $\alpha$ on sound speed did not receive proper
attention.  Although protoplanetary disks
are thin, this rise in angular momentum transport with temperature suggests
that the effects of radiation will be important for high temperature regions
around forming planets as well as  regions where the turbulence leads to
significant Joule and/or viscous heating. 

\subsection{Excluding the inner boundary}

The correct treatment of boundaries is a major issue for numerical
simulations. In global simulations of disks, outflow or frozen boundaries are
usually used, both of them being more realistic than reflective, but also
presenting disadvantages. Strictly speaking, a ``perfect'' boundary might well
not exist. The best solution would be, of course, not having to use a boundary
at all.

Cylindrical grids only make sense if the center of mass is at the origin of
the coordinate system as only then angular momentum transport can be written in a conservative form. If the center of mass is not at the origin or there is
more than one massive object the cylindrical coordinate system artificially
introduces a reflective boundary at the center. Cartesian grids, as stated
before, are not hindered by this, and we can therefore study how the presence
or absence of an inner boundary affects the results.

We compute a version of model A where the computational domain extends all the
way to \rint=0. Without an inner boundary, the gravitational potential, the
angular frequency and the sound speed have to be smoothed according to
Eq.~(\ref{smoothing}) to prevent singularities. When taking global averages, we
exclude the smoothed region.

The evolution of the turbulence and the globally averaged stresses at
saturation are almost identical to those seen at model A. The only noticeable
difference is that the highly fluctuating magnetic field observed in the inner
boundary of model A (with rms amplitude $\approx$ 2 times seen in the rest of
the disk) does not occur in this model. 

In model A, the magnetic potential at the inner boundary
remains frozen at the initial condition $\vv{A}$=$0$. As the MRI builds up
the magnetic potential in the freely evolving disk, a sharp radial gradient
appears at the boundaries. This sharp radial gradient in the azimuthal and
vertical components of the magnetic potential translates into high values of
the magnetic field. If we were solving for the magnetic field instead, a
similar effect would be seen. The magnetic field would rise in the disk, but is kept frozen at the boundary, thus building up a high magnetic pressure with equally damaging effects for the simulation.

By avoiding the inner boundary altogether, such an effect does not occur.
However, other problems arise as the advection on the very inner disk
($s$$<$0.4) happens on tight circular trajectories that are poorly resolved in
the Cartesian geometry near the center of the grid. As they behave like highly
localized vortices, in most cases the explicit shock dissipation terms
ensure numerical stability. But for cases with stronger turbulence (model D,
Sec 4.4), a model without an inner boundary could not be treated.

As we did for the models with an inner boundary, we check the dependency of the
turbulent stresses with sound speed by computing versions of models B and C
without an inner boundary. As for model A2, these models B2 and C2 behave quite
similarly to models B and C, saturating at roughly the same stresses. 

\subsection{Radially varying field - models D and E}

With the constant vertical field, it is seen that the inner disk does not go
turbulent. This is due to the fact that in this rapidly advecting region, the
growth of the MRI is numerically damped.

In view of this, we also compute models with a radially varying $z$-field 
\begin{equation}
\label{radial-field}
  \vv{B} = \frac{L_z \Omega(s)}{2  j \pi}\sqrt{\mu_0\rho_0} \hatz, 
\end{equation}
such that $j$ Balbus-Hawley wavelengths (Eq.~[\ref{balbus-hawley}]) are
resolved through the vertical extent $L_z$ of the box at any radius. We
computed a model with $j=4$ (model D), and another version, with a
weaker field (model Dw) with $j=16$, so that unstable wavelengths are 
reasonably well resolved. These fields can make the whole disk go
turbulent, but they grow so strong in the inner disk of model D that a more
pronounced temperature profile ($q_{_T}$=$2$) had to be used to avoid the
magnetic field from going superthermal. For model D, such a setup has plasma
$\beta$=20 at \rint  and 120 at \rext. Model Dw has a field with a strength
more similar to model A, and therefore would not need this fix. However, to
allow a comparison with model D, we also used this steeper temperature
gradient, thus having a plasma $\beta$= 300 at \rint  and 1850 at \rext.

As this setup has an inverse $\beta$ profile compared to the models
with a
constant vertical field, the turbulence starts from the inner disk instead of
the outer. Also, as the magnetic field is stronger, wavelengths of faster 
growth rate are better resolved and the turbulence saturates after a few 
orbits. Due to the strong stresses, we had to raise shock resistivity to 2 
instead of 1 as used before.

\begin{figure*}
\begin{center}
\resizebox{14.75cm}{!}{\includegraphics{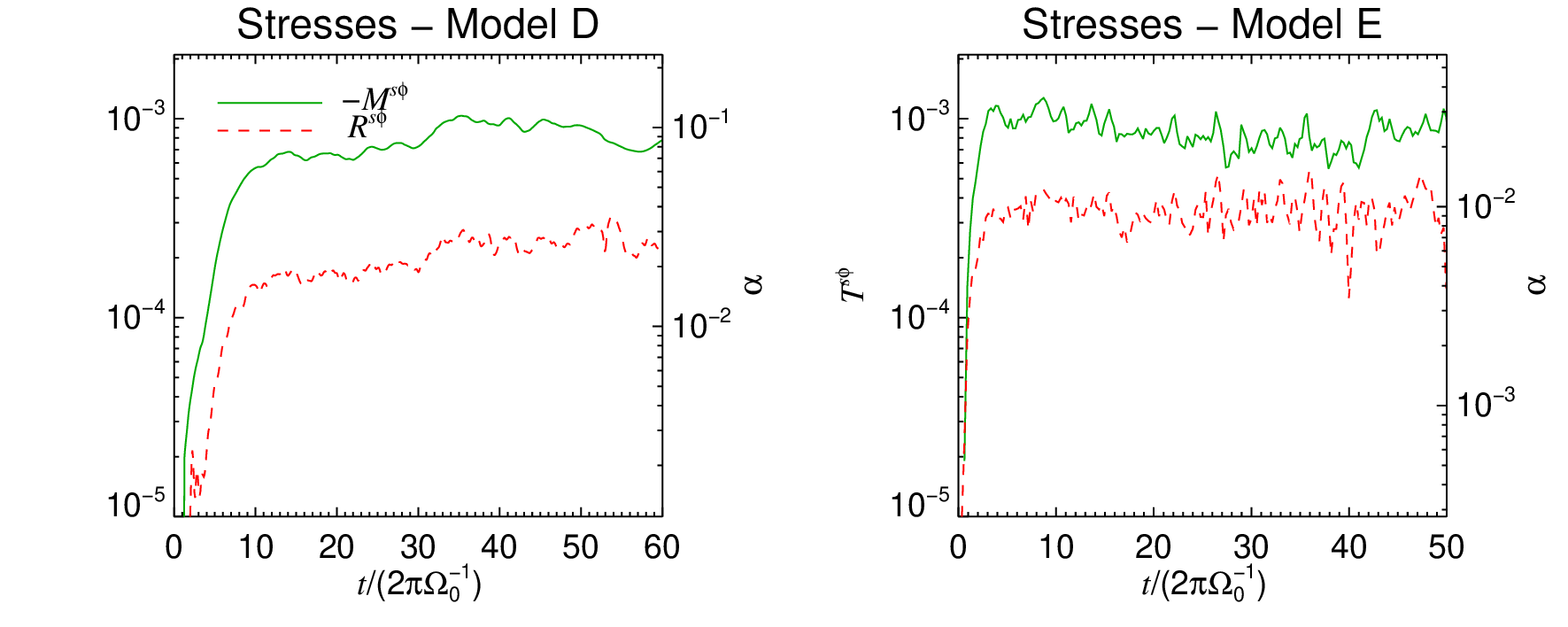}}
\resizebox{14.75cm}{!}{\includegraphics{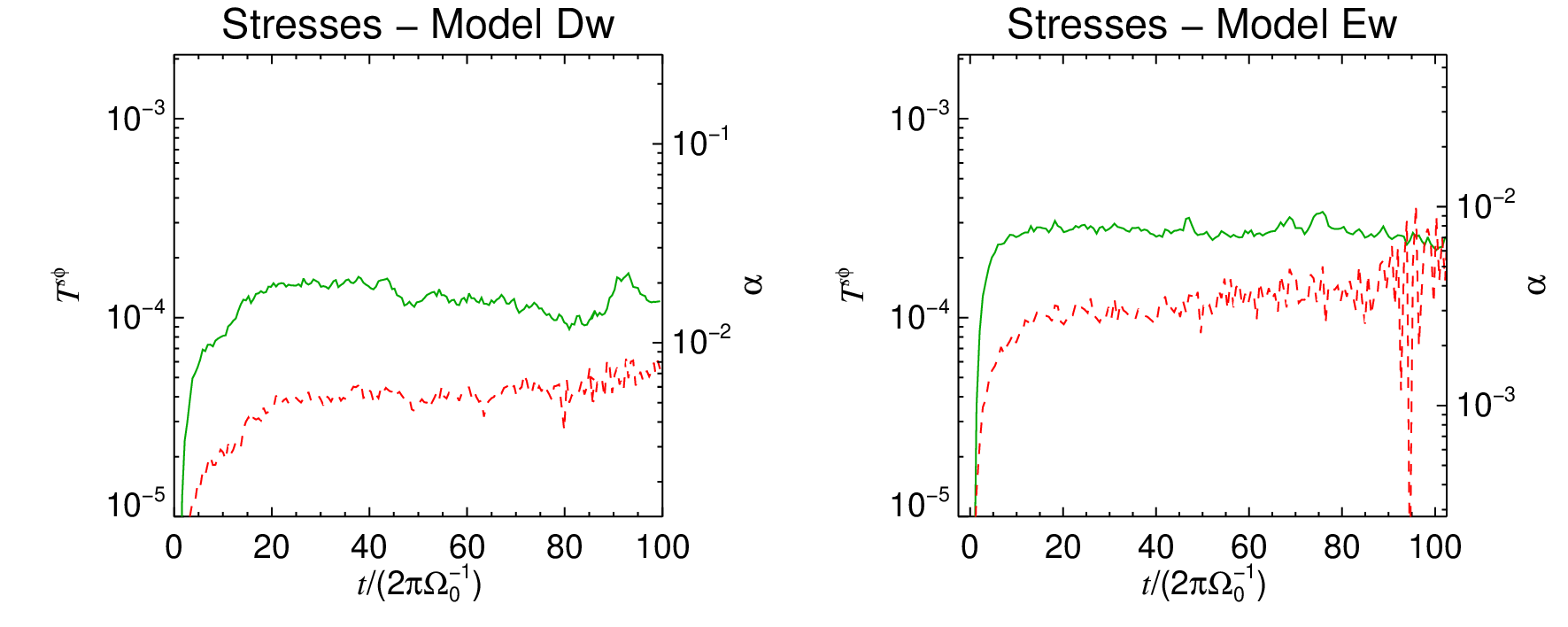}}
\end{center}
\caption[]{Time evolution of the turbulent stresses for models D and E
(upper panels) and Dw and Ew (lower panels). As compared with model A, the
stresses saturate at a much earlier time. The ratio of Maxwell to Reynolds
stress is around three for model D, and around two for the hotter model E,
approaching one at the end of the simulation. Notice that as seen in models
ABC, the stresses are bigger for the hotter model, although the alpha viscosity
value is smaller. Time is quoted in orbits at $s_0$.}
\label{radial-field-stress}
\end{figure*}

\begin{figure*}
\begin{center}
\resizebox{8cm}{!}{\includegraphics{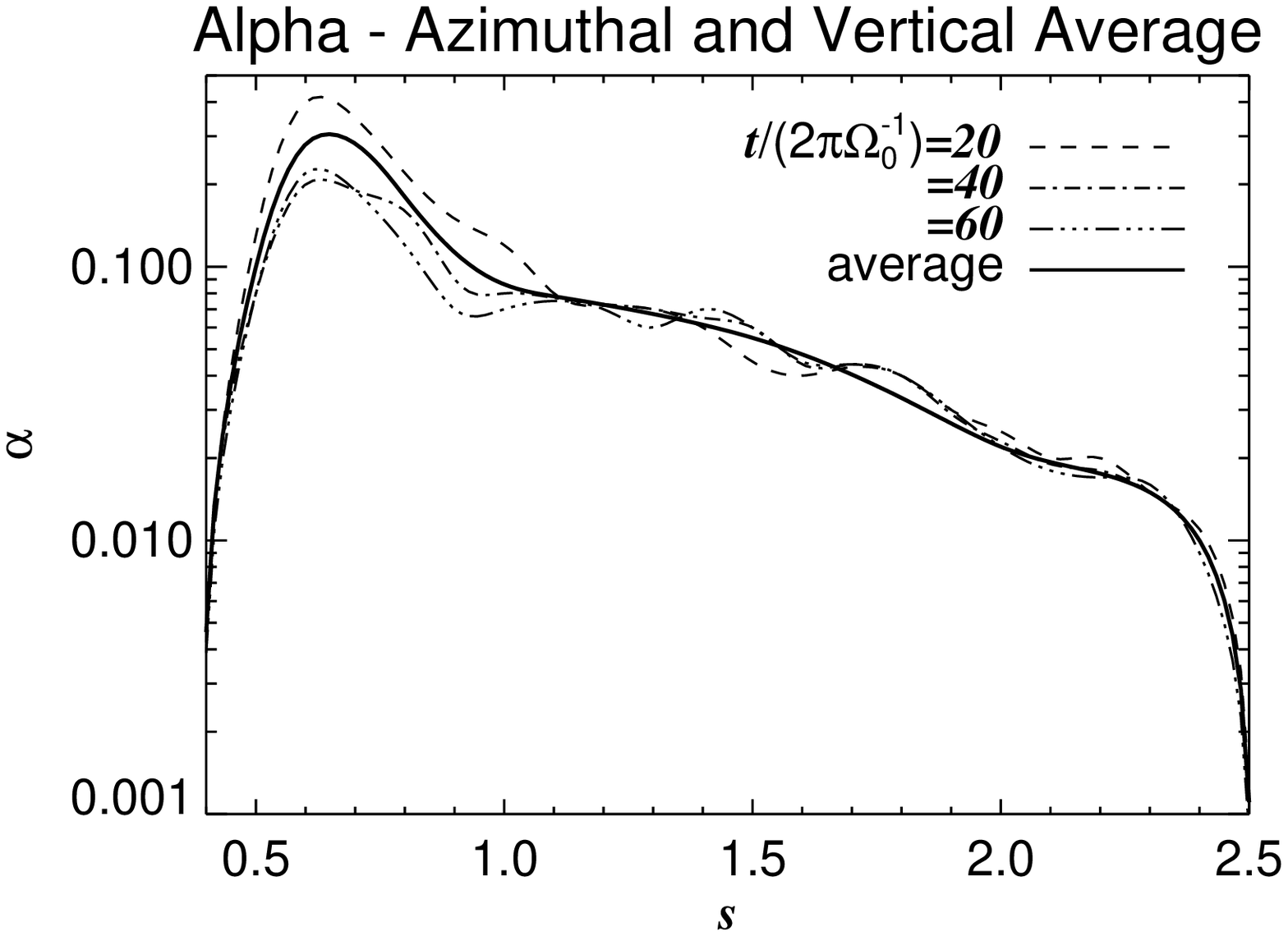}}
\resizebox{8cm}{!}{\includegraphics{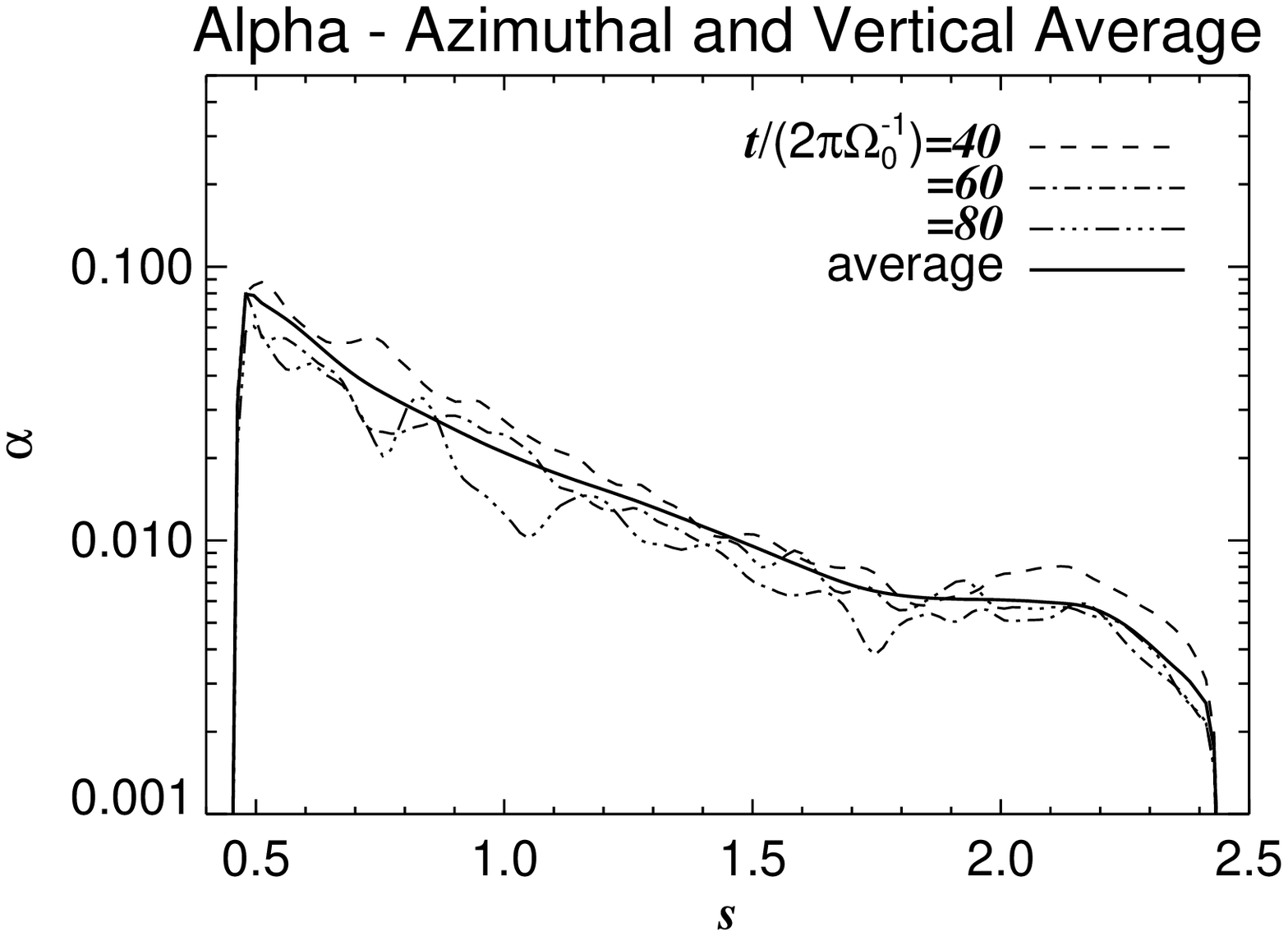}}
\end{center}
\caption[]{Same as Fig.~\ref{radalphacs5}, but for models D (left panel) and Dw
(right panel). The times corresponding to the snapshots are indicated in the
legends, as well as the time average. The inner disk is considerably more
turbulent than the outer parts.} 
\label{radalpha-D}
\end{figure*}

The stresses also saturate at higher values, 0.08 for the Maxwell stress, and
0.02 for the Reynolds stress (Fig.~\ref{radial-field-stress}) for model D.
The weaker field yields a total alpha viscosity around $\xtimes{2}{-2}$
($\alpha_M$=0.017 , $\alpha_R$=0.005, see Fig.~\ref{radial-field-stress}a and
Table~2). These values are at least one order of magnitude higher than the
ones obtained with a constant field. The radial structure of the alpha value,
plotted in Fig.~\ref{radalpha-D}, reveals that the stresses follow the radial
profile of 1/$\beta$, being stronger in the inner disk. The same was seen in
the simulations with a constant field, where the stresses were stronger in the
more magnetic outer disk.

The stresses in this model are so high that the alpha viscosity in the 
saturated state is always of the order $\ttimes{-1}$, reaching $\approx$0.5 in the more
magnetized inner disk (Fig.~\ref{radalpha-D}). That means that the inner disk
approaches magnetically dominated values for plasma $\beta$ as the turbulence
saturates. The global average of plasma $\beta$ in the saturated state is 4,
although it did not reach superthermal values ($<$1) during the course of the
simulation. Model Dw is milder, having a total alpha value of the order
$\ttimes{-2}$, reaching a maximum of $\approx$0.08 in the more magnetized inner
disk.

The high stresses compared to the cases with constant vertical field stem from the
initially stronger magnetic field, not from a pressure effect. Indeed, in this
setup, the temperature profile is steeper, but it never rises above the sound
speed of model C, with $c_{s_0}=0.2$ and power law of exponent $0.5$. However,
we still expect the pressure effect seen on models ABC to be present. To check
the behavior of this setup with the imposed temperature profile, we compute
another model (model E) with the same initial condition for the magnetic field 
as model D, but with a hotter temperature, with $c_{s_0}=0.2$. This setup has
a plasma beta value of 80 in the inner disk and 450 in the outer. A weaker
version with plasma beta value of 1200 in the inner disk and 7400
(model Ew) was also computed, for comparison with model Dw. The results
(Fig~\ref{radial-field-stress}b) are similar to models ABC: The stresses are
indeed larger than those of model D and Dw, but the alpha viscosity values are
smaller.  Also as seen on models ABC, the kinetic alpha did not change
appreciably, but the magnetic alpha was reduced by approximately a factor of 2.

The ratio of stresses varied as well, as seen in models ABC. It is around
4 for model D, but 2 or less for model E. It is around 3 for model Dw, but less
than 2 for model Ew. 

A final note on the temperature effect; we see that, like alpha, the
turbulent plasma beta parameter $\beta_t$ does not scale with pressure. As the
rise in the turbulent magnetic energy is outpaced by the growth of the
pressure, $\beta_t$ increases with increasing temperature.
 
\begin{figure*}
\begin{center}
\resizebox{15.25cm}{!}{\includegraphics[angle=90]{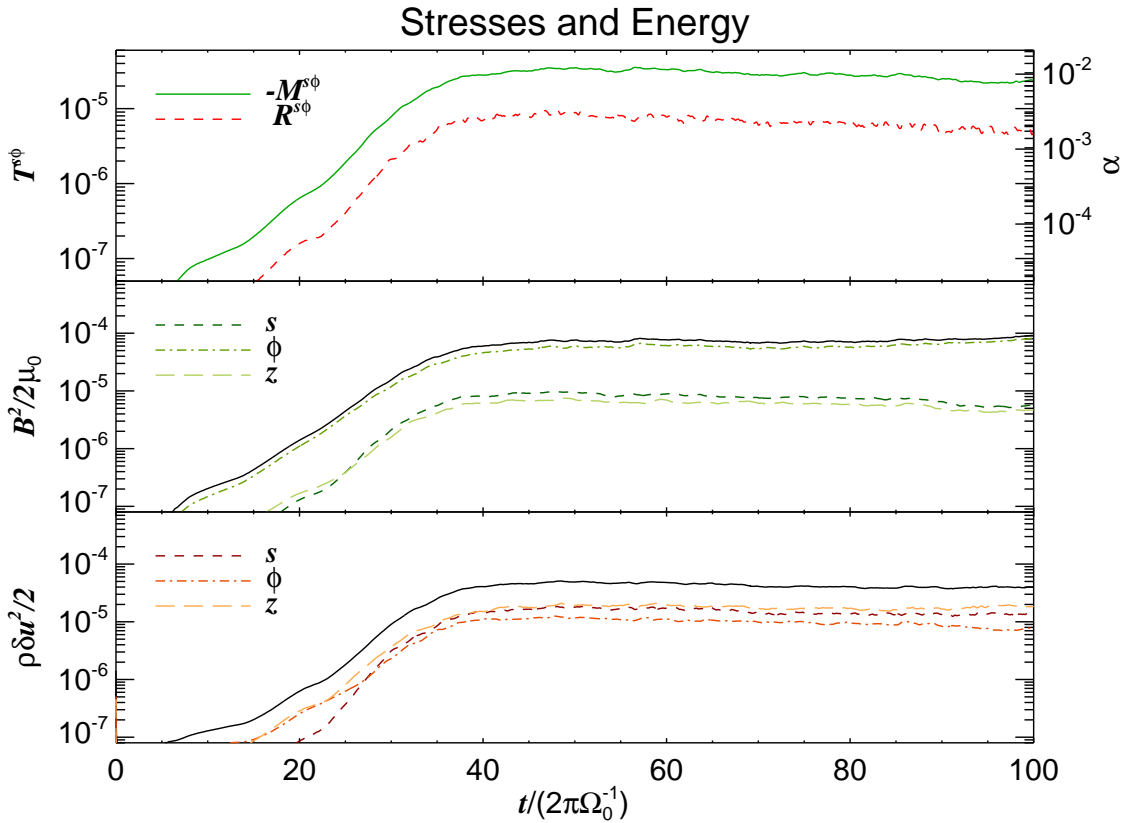}}
\end{center}
\caption[]{Same as Fig.~\ref{fiducialcyl} (constant vertical field), but for
model F (constant azimuthal field).} 
\label{modelF-stress-energy}
\end{figure*}

\begin{figure*}
\begin{center}
\resizebox{15.25cm}{!}{\includegraphics{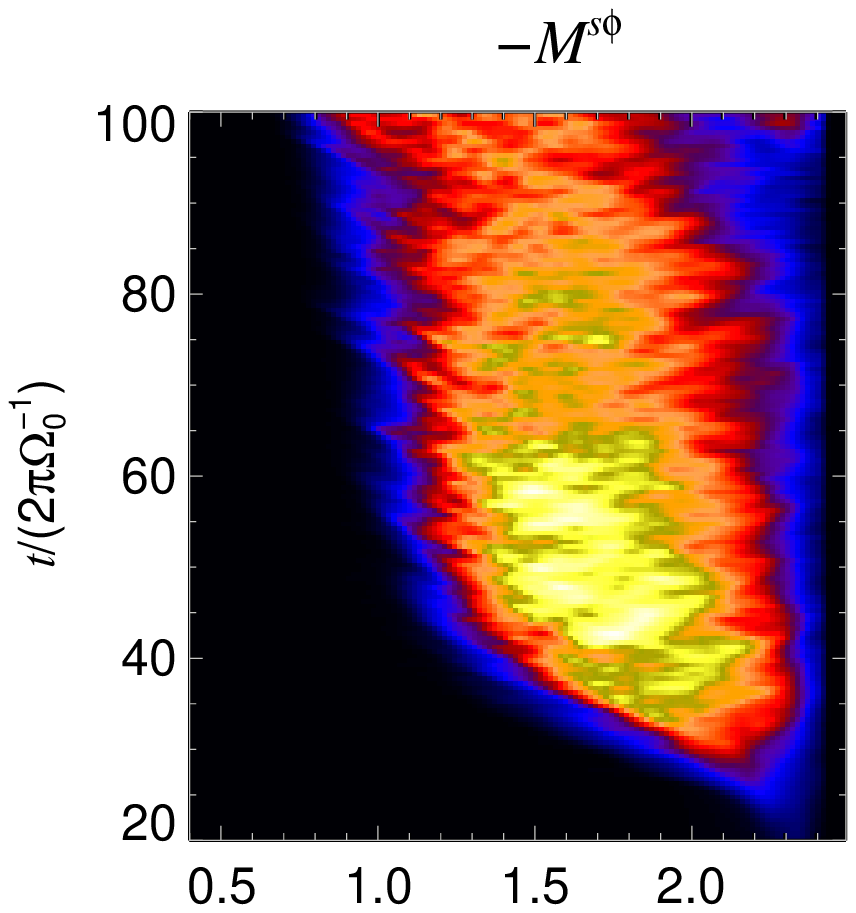}}
\resizebox{15.25cm}{!}{\includegraphics{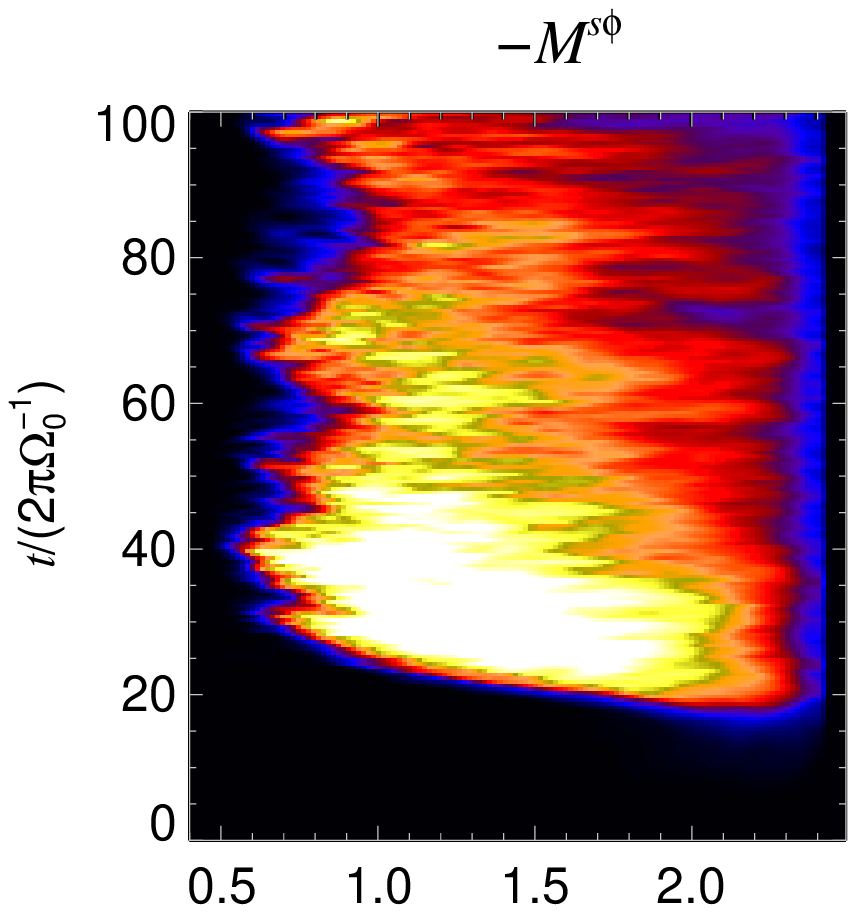}}
\end{center}
\caption[]{Space-time diagram of the turbulent stresses and alpha
viscosity for
model F (upper panels) and model G (lower panels). Time is quoted in 
orbits at $s_0$=1.0. Saturation is reached at 40 orbits, but after 60 orbits 
the stresses seem to start a slow decay. The alpha
viscosity appears constant, due to a similar decay in gas pressure in the outer
disk, that starts to deplete as the resulting accretion builds a negative
density gradient. Model G behaves similarly, but with higher stresses, 
lower alpha viscosity and being turbulent further inside.}
\label{modelF-alpha}
\end{figure*}

\subsection{Constant azimuthal field - models F and G}

Analytical treatment (Balbus \& Hawley 1992, Ogilvie \& Pringle 1996) and
numerical simulations (Hawley 2000, Papaloizou \& Nelson 2003) show a wealth of
evidence that the MRI also exists for a purely toroidal field. In this case, 
waves of the form $\exp[i (m\phi - \omega t)]$, where $m$=$k_\phi
s$ is the azimuthal wavenumber, are excited. The maximum growth rates are
similar to those observed in purely vertical fields, but reached at much
smaller azimuthal wavenumbers.

For an azimuthal field, the maximum growth rate occurs at the wavenumber
(Balbus \& Hawley 1998)
\begin{equation}
  m_{\small{\rm max}}(s)=\frac{\sqrt{15}}{2} \frac{\Omega s}{\va}
\end{equation}
Such wavelengths are now resolved in the $xy$ plane, instead of in the vertical
direction. Without the severe constraint of fitting unstable wavelengths in the tiny vertical scale height 
of the disk, the azimuthal field can be set at much stronger values than those used in the vertical cases. The only constraint is that we keep the field subthermal. With a temperature gradient of $q_{_T}$=$1$ and  
$c_{s_0}$=0.05, a constant azimuthal field of $B_0$=$\xtimes{3}{-2}$ (model F) corresponds to plasma 
$\beta$ of 12 at $s$=0.4, 5.5 at $s_0$ and 2 at
$s$=2.5. The wavenumbers are $m_{\small{\rm max}}$=65 at $s_0$, 102 at $s$=0.4 and 41 at the outer
boundary. A hotter version, with $c_{s_0}$=0.20, yielding plasma
$\beta$=220 at $s$=0.4, 90 at $s_0$ and 35 at $s$=2.5 (model G), was also
computed for comparison.

Following the time evolution of model F, we see that the turbulence actually saturates at 
$t$=200 at $s$=2.0 ($\approx$ 11 orbits), $t$=300 at $s$=1.5
($\approx$ 30 orbits) and is still growing linearly at $s$=1.0 at the end of the
simulation at $t$=628 (100 orbits). The global average
(fig.~\ref{modelF-stress-energy}), and a space-time $(s,t)$ inspection
(fig.~\ref{modelF-alpha}) of the stresses reveals that after reaching
saturation at $t$=250 (40 orbits at $s_0$), a steady state is maintained for 20
orbits, after which the turbulence starts decaying slowly. But as a small
growth is observed near the end of the simulation, it is not clear if this
decay would continue to zero or if it constitutes just a unusually long
fluctuation of the turbulence. Moreover, the magnetic and kinetic energies
(fig.~\ref{modelF-stress-energy}) show no signs of decaying.

On the right panel of fig.~\ref{modelF-alpha}, we plot the total alpha
viscosity parameter $\alpha_M + \alpha_R$. Curiously, it does not show the
decaying effect seen on the stresses, implying that a decrease in gas
pressure accompanied the decrease in Maxwell stress. Such a behavior is
expected, since a negative density gradient is arising from the accretion
process. Therefore, a depleted outer disk has a larger value of alpha viscosity
for the same Maxwell stress.

The global average yields $R^{s\phi}$=(0.7$\pm$0.1)$\times\ttimes{-5}$,
$M^{s\phi}$=(2.9$\pm$0.4)$\times\ttimes{-5}$ and total alpha viscosity
$\alpha$=(1.3$\pm$0.1)$\times\ttimes{-2}$.

Model G shows Maxwell stresses that go further towards the inner disk, 
and Reynolds stresses that extend to inside $s$=$0.4$ as 
seen in Fig.~\ref{modelF-alpha}. The outer disk attains a turbulent state 
first, at $t$=100 (15 orbits), and the inner disk at $t$=200 (30 orbits). The turbulent alpha
parameter peaks at $\xtimes{8}{-3}$ ($\alpha_M$=$\xtimes{6}{-3}$,
$\alpha_R$=$\xtimes{2}{-3}$) at 30 orbits and starts a long decay until
leveling after other 30 orbits at $\alpha_M$=$\xtimes{3}{-3}$ and
$\alpha_R$=$\xtimes{7.5}{-4}$. The global averages are
$R^{s\phi}$=(5$\pm$2)$\times\ttimes{-5}$,
$M^{s\phi}$=(1.8$\pm$0.6)$\times\ttimes{-4}$ and total alpha viscosity
$\alpha$=(4.2$\pm$1.5)$\times\ttimes{-3}$.

The referee, Dr.\ Ulf Torkelsson, pointed out to us that the non-zero
tension of the constant azimuthal field,
\begin{equation}
  \mu^{-1}(\vv{B}\cdot\del)\vv{B} = -\mu^{-1}B_0^2 s^{-1} \, \hats \, , \nonumber
\end{equation}
leads to an increase in the centripetal force that is not taken into account by 
Eq.~(\ref{MHStatic-equilibrium}), so the models with azimuthal flux are not started in strict 
magnetohydrostatical equilibrium. We therefore performed a set of 2D tests 
to assess how this out-of-equilibrium initial 
condition could modify our results. First we ran a 2D disk without noise in 
the velocity field, so although a magnetic field is present, the spectrum of 
wavelengths is not excited. The departure from equilibrium launches a sound 
wave starting from the inner disk and propagating outwards. At time $t$=30 
($\approx$5 orbits) in model F the sound wave reaches the outer boundary and 
is damped by the buffer zone. After that, the oscillations slowly damp 
through the next orbits as the disk settles into centrifugal equilibrium 
between gravity, thermal pressure, and magnetic tension forces. We followed 
the evolution until time $t$=90. At this time, the amplitude of the perturbation dropped 
to 3\% of the initial density and 1\% of the reference sound speed $c_{s0}$. 

In model G, with a sound speed 4 times faster, the sound wave reaches the 
outer boundary much earlier, at time $t$=12 (about 2 orbits). By including 
noise we see the same results, so we conclude that in a 2D case, even though 
the non-vanishing magnetic tension leads to an out of equilibrium initial 
configuration, the discrepancy is slight and the system quickly relaxes in a 
timescale that is much smaller than the time the MRI takes 
to saturate (20 orbits). 

\section{Disks with solid boulders}

Having presented the gaseous disk models, we now proceed to study the behavior
of solid boulders inserted in these disks we have constructed. Meter-sized
boulders are an important step towards kilometer-sized planetesimals. They are
also interesting from a gas-dynamical point of view because they are only
marginally coupled to the gas (on approximately a Keplerian shear time-scale)
and can thus experience concentrations in vortices and transient gas high
pressures (Barge \& Sommeria 1995, Fromang \& Nelson 2005, Johansen et al. 2006)

Our models are usually evolved for $\approx$75 orbits at $s_0$ before we add
the particles, to allow for the turbulence to develop and saturate. A large
number of particles ($\ttimes{6}$) is used, which allows us to trace the swarm
of particles onto the grid as a density field. The initial condition is such
that the particles are concentrated in an annulus of constant bulk density
$\rho_p$, ranging from \rint to \rext, with a solids-to-gas ratio of 0.01,
typical of the interstellar medium. Their velocity is initially the Keplerian
angular velocity for their radial location. 

As boundary conditions, we do not allow the particles to leave the
computational domain as they drift inwards due to gas drag from the slightly
sub-Keplerian gas. Instead, a particle that crosses the inner radius \rint will
be relocated to the outer radius \rext, where it will reappear at the same
azimuthal location, thus mimicking periodic boundary conditions. Its velocity,
however, will be reset to the Keplerian angular velocity at \rext. A particle
that tries to cross the outer radius will simply have its velocity reset to
Keplerian without changing its radial or azimuthal location. 

We include particles at the later stages of models A and D. For model A, where
the inner disk does not go turbulent, it was noticed that by allowing the
particles to move through all the radial range, they eventually got trapped in
the several local density maxima between the concentric rings that the inner
disk breaks into. As such a loss of particles is undesirable, we keep them
where the turbulence is saturated by setting the inner radius of the boundary
conditions for particles outwards of $s_0$. In model D, where the Balbus-Hawley
wavelength is resolved at all radii, such correction is not needed. We discuss
the results of model D first.

\subsection{Particles on model D}

\begin{figure*}
\begin{center}
\resizebox{15.25cm}{!}{\includegraphics{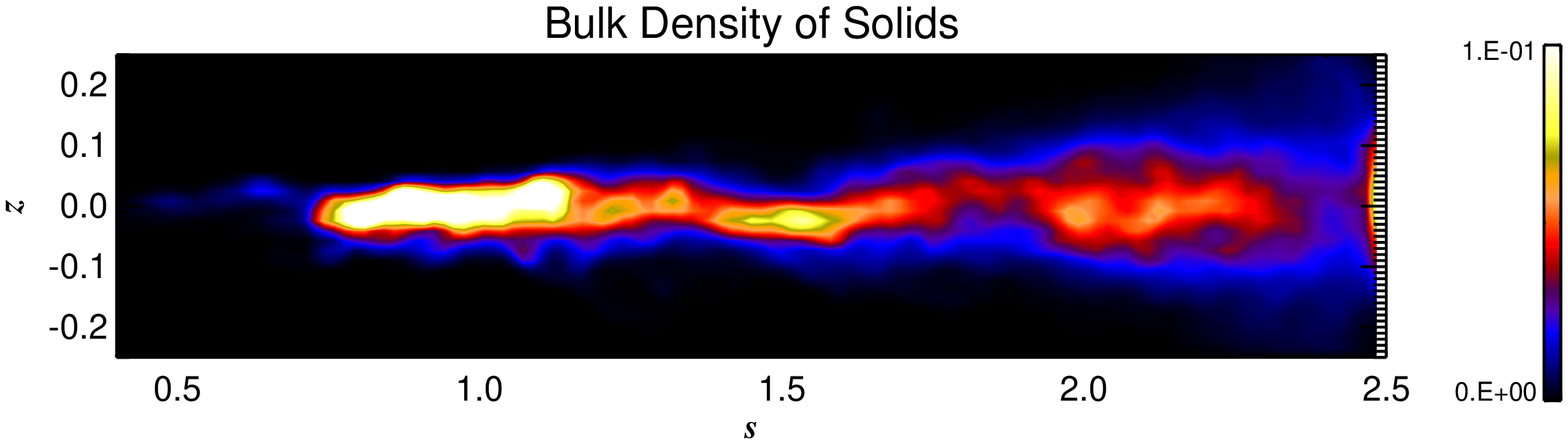}}
\resizebox{15.25cm}{!}{\includegraphics{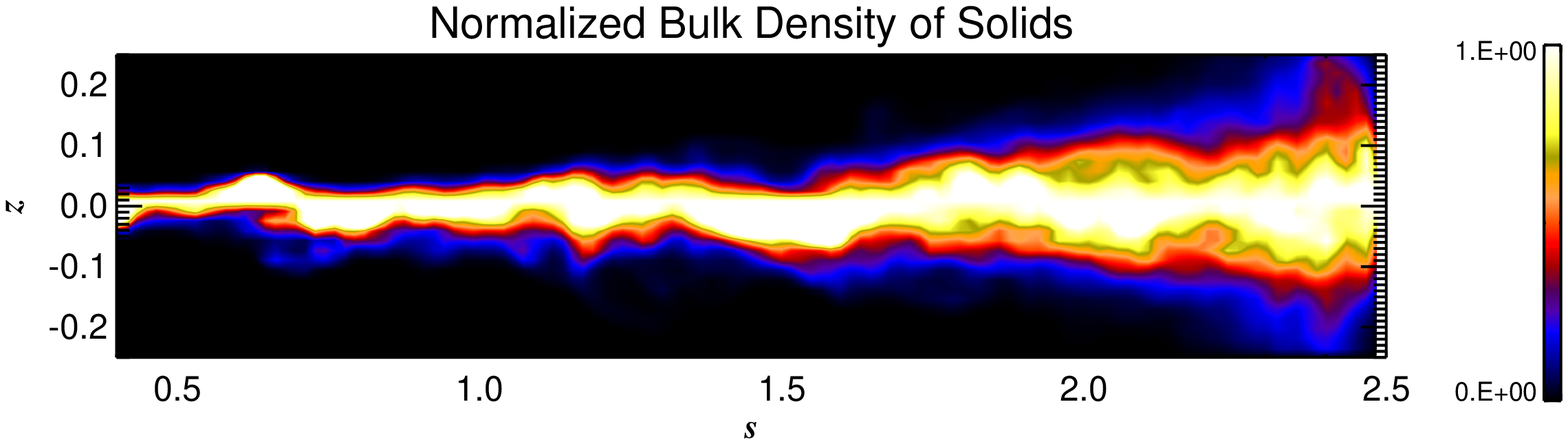}}
\end{center}
\caption[]{Vertical slice of the bulk density of solid particles (upper) and
the same quantity normalized by the midplane density. A midplane layer forms in
equilibrium between sedimentation and turbulent diffusion. The scale height of
this layer follows a linear dependence with radius.}
\label{dust-density}
\end{figure*}

The particles soon fall to the disk midplane due to the vertical gravity. At
the same time they get trapped in high pressure regions in the turbulent flow
due to the drag force (Klahr \& Lin 2001, Johansen et al. 2006).  In
Fig.~\ref{dust-density} we plot a slice of the bulk density of solids profile
10 orbits after the insertion of the particles into the simulation. A
pile-up of solids in the inner disk is seen to have occurred, because particles
have concentrated in a pressure maximum in the gas (we discuss this further in
Sect.\ \ref{particles_a}; see also Fig.\ \ref{midplanes}).

As discussed by Johansen \& Klahr (2005), while solid particles are pulled
towards the midplane by the stellar gravity, turbulent motions stir them up
again. A sedimentary layer in equilibrium between turbulent diffusion and
gravitational settling is formed. The thickness of this layer is therefore a
measurement of the turbulent diffusion acting on the solid particles. 

Under the influence of gravity, the solids settle with a profile similar to the
one generated by a pressure force (Dubrulle et al. 1995)
\begin{equation}
\label{dust-scale}
  \ln \rho_{\rm p}(s,z) = \ln \rho_{\rm p}(s,z=0) - \frac{z^2}{2H_{\rm p}^2}.
\end{equation}

By comparing this profile with the
analytical expression for a pressureless fluid under diffusion, gas drag and
vertical gravity (Johansen \& Klahr, 2005) 
\begin{equation}
\label{analytical-anders}
  \ln \rho_{\rm p} = \ln \rho_{\rm p}(s,z=0) - \frac{\tau_f}{D_z^{(t)}} \int g_z\, {\de}z,  
\end{equation}
and recalling that $g_z=-\Omega^2 z$, we have 
\begin{equation}
\label{turb-diffusion}
  D_z^{(t)} = \Omega^2 H_{\rm p}^2 \tau_f 
\end{equation}
From Eq.~(\ref{dust-scale}), we see that the scale height of the solids is the
vertical distance in which the bulk density falls by a factor $1/\sqrt{\ee}
\approx 0.6$ relative to the value at midplane. We plot in
Fig.~\ref{dust-density}b the bulk density normalized by its value in the
midplane. In this figure, the quantity plotted is in fact identical to the
exponential term in Eq.~(\ref{dust-scale}). Where it reaches 0.6, the vertical
distance $z$ gives the diffusion scale height $H_{\rm p}$. 

We fit the points where the exponential term equals 0.6 with a power law
$H_{\rm p}=a r^{n}$. A linear regression in logarithm yields $a=0.042$ and
$n=0.97$, with an rms of 0.04. This translates into a diffusion coefficient
(Eq.~[\ref{turb-diffusion}]) of $D^{(t)}\approx \xtimes{1.7}{-3}\,s^{-1}$. As
this model has a sound speed profile $c_s$=$0.1\,s^{-1}$, the diffusion
coefficient in dimensionless units corresponds to
$\delta^{(t)}=D^{(t)}c_s^{-2}\Omega=0.17\,s^{-1.5}$, or 0.14 if globally
averaged. The rms of 0.04 in the logarithm fit yields an uncertainty of 0.01 in
this global average. As the total alpha viscosity is 0.112$\pm$0.003, the
globally averaged vertical Schmidt number, i.e., the strength of viscosity when
compared to vertical diffusion, is 0.78$\pm$0.06. 

\subsection{Particles in model A}
\label{particles_a}

For model A, the alpha viscosity ($\alpha \sim \ttimes{-3}$) is much lower than
in model D ($\alpha \sim \ttimes{-1}$), so according to
Eq.~(\ref{turb-diffusion}) and assuming that the diffusion coefficient is of
the same order of the turbulent viscosity, we expect the sedimentary layer of
solid particles to have a scale height of $H_{\rm
p}=\alpha/\Omega^2\tau_f\approx0.03 H$. At \rext, the gas scale height $H$
equals 0.08 for $c_{s_0}=0.05$, then $H_{\rm p}=\xtimes{2.5}{-3}$. With a grid
resolution $\Delta z=0.02$, this layer will not be resolved. It means that the
interpolation of the particle density back to the grid will not allow for a
grid-based measurement of the diffusion acting in the turbulent layer as we did
for model D. 

\begin{figure*}
\begin{center}
\resizebox{15.25cm}{!}{\includegraphics{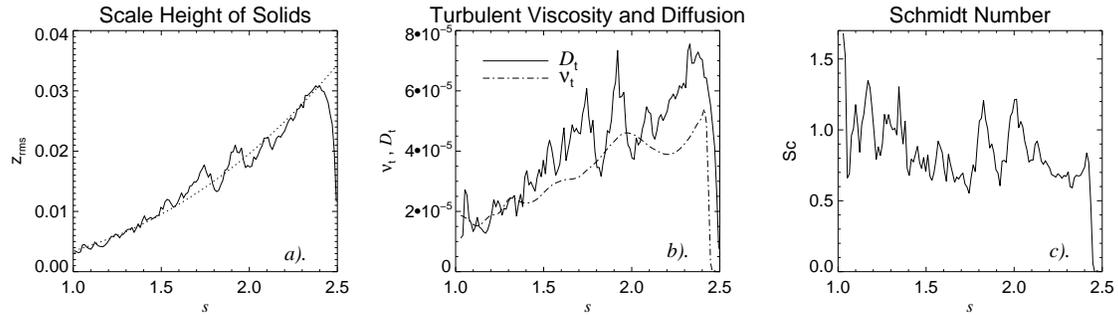}}
\end{center}
\caption[]{Time-averages of the dispersion of vertical positions of particles
w.r.t.\ the disk midplane, revealing the scale height of the sedimentary layer
that forms due to the equilibrium between sedimentation and turbulent diffusion
for model A. The dotted line is a power law fit, yielding an exponent of
$\approx$ 2.5. This radial profile is a time average between orbits 4 and 20
(see text).  The resulting diffusion coefficient resulting from this scale
height is shown as solid line in the middle panel. The time averaged gas-phase
viscosity is shown in dot-dashed line for comparison. It is seen that these two
coefficients have similar strength. The right panel shows the Schmidt number of
the flow. It is indeed close to unity through the radial domain.}
\label{dust-scale-height}
\end{figure*}

But as the particles are Lagrangian, we can plot their real positions and trace
the scale height of solids in a grid-independent way. The diffusion process
operates in much the same way, nearly independent of grid resolution, since the
large scale velocities of the gas are well resolved. In order to do this, we
define 128 bins in the radial direction and measure the individual vertical
positions of the swarm of particles with respect to the midplane of the disk
within these bins. The standard deviation $z_{\rm rms}$ of the vertical
positions of particles with respect to the midplane in each bin immediately
gives the scale height of the sedimentary layer. The result of this process is
shown in Fig.~\ref{dust-scale-height}a, where we average $z_{\rm rms}$ as
measured on 17 snapshots, from orbits 4 to 20 at $s_0$ after the insertion of
the particles.  The initial time is chosen at 4 orbits because it is the time
it takes for the drag force to couple the particles to the gas at \rext, thus making sure that the sedimentary layer is in equilibrium between
gravitational settling and turbulent diffusion through the whole radial extent
of the disk.

The dashed line in Fig.~\ref{dust-scale-height}a represents the power law fit
$H_{\rm p}$=$a r^{n}$ to the measured scale height. It yields $a$=$0.003$ and
$n$=$2.48$. The rms of the logarithmic fit is $0.09$. In
Fig.~\ref{dust-scale-height}b we plot the resulting diffusion coefficient
$D_z^{(t)}$=$ \Omega^2 H_{\rm p}^2 \tau_f$. It can be approximated by a power
law of $\approx \xtimes{9}{-6}\,s^{2}$. In dimensionless units it corresponds
to $\delta^{(t)}$=$\xtimes{3.6}{-3}\,s^{-1.5}$, or 0.007 if globally averaged.
The uncertainty is 0.001. This behavior of the turbulent diffusion that acts on
solids is quite similar to the one shown by the turbulent viscosity arising
from the stresses on the gas phase (dot-dashed line in
Fig.~\ref{dust-scale-height}b). Checking the radial dependency of the Schmidt
number (Fig.~\ref{dust-scale-height}c), we see that it is of the order of unity
all over the radial domain. A slight trend is seen towards smaller Schmidt
numbers in the outer disk, but it never gets below 0.6.

\begin{figure*}
\begin{center}
\resizebox{15.25cm}{!}{\includegraphics{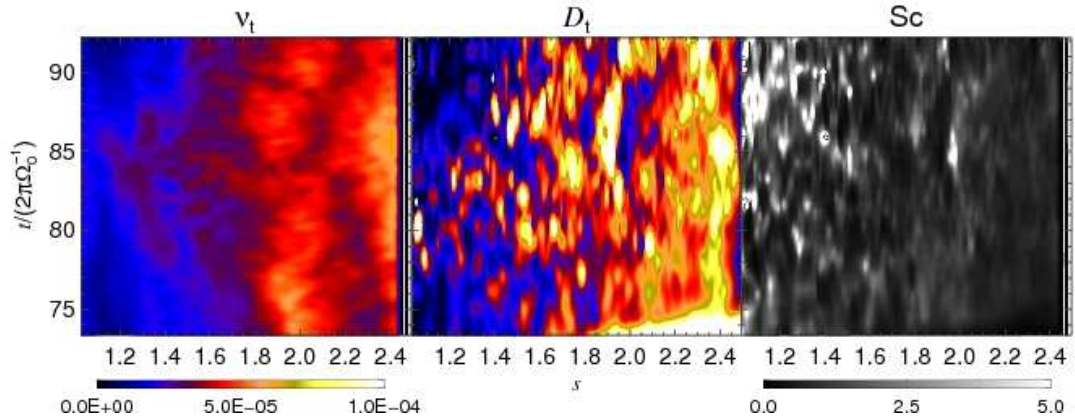}}
\end{center}
\caption[]{Azimuthally and vertically averaged turbulent viscosity $\nu_t$,
turbulent diffusion $D_t$ and the resulting Schmidt number Sc as a function of
radial position and time for model A. Time is quoted in orbits at $s_0$.
Viscosity and diffusion are shown in the same units and color-code. Some
localized overdensities in diffusion last for some orbits, while viscosity
shows a smoother evolution in time. The quantities have approximately the same
strength, as the Schmidt number is overall around unity, and seldom greater
than 5.}
\label{spacetime-schmidt}
\end{figure*}

In Fig.~\ref{spacetime-schmidt} we explore this radial dependency in more
detail. The figure shows the vertical and azimuthal average of the turbulent
viscosity $\nu_t$, turbulent diffusion $D_t$ and their ratio Sc, the Schmidt
number, as a function of radial position $s$. On the time axis we show time in
orbits at $s_0$ since the beginning of the simulations. At 71 orbits the
particles are inserted, and quickly fall to the midplane. As seen from the
diffusion map, the particles in the innermost radii quickly sediment, settling
in a diffusive equilibrium in less than two orbits. After four orbits, $t=75$,
the last radius achieves diffusion equilibrium and the situation becomes
statistically unchanged until the end of the simulation.

The gas-phase viscosity and the solid-phase diffusion are similar in average,
but the diffusion is seen to fluctuate more. Transient patches of high
diffusivity are seen to live for some orbits at a constant radial location
before decaying. The turbulent viscosity, in turn, appears much smoother in
time. The resulting Schmidt number is shown in greyscale in the right panel of
Fig.~\ref{spacetime-schmidt}. Due to the high variability of diffusion, some
short lived bright areas of Sc$>$5 are seen, but overall the Schmidt number is
around 1 throughout the space-time domain.

\begin{figure*}
\begin{center}
\resizebox{ 7cm}{!}{\includegraphics{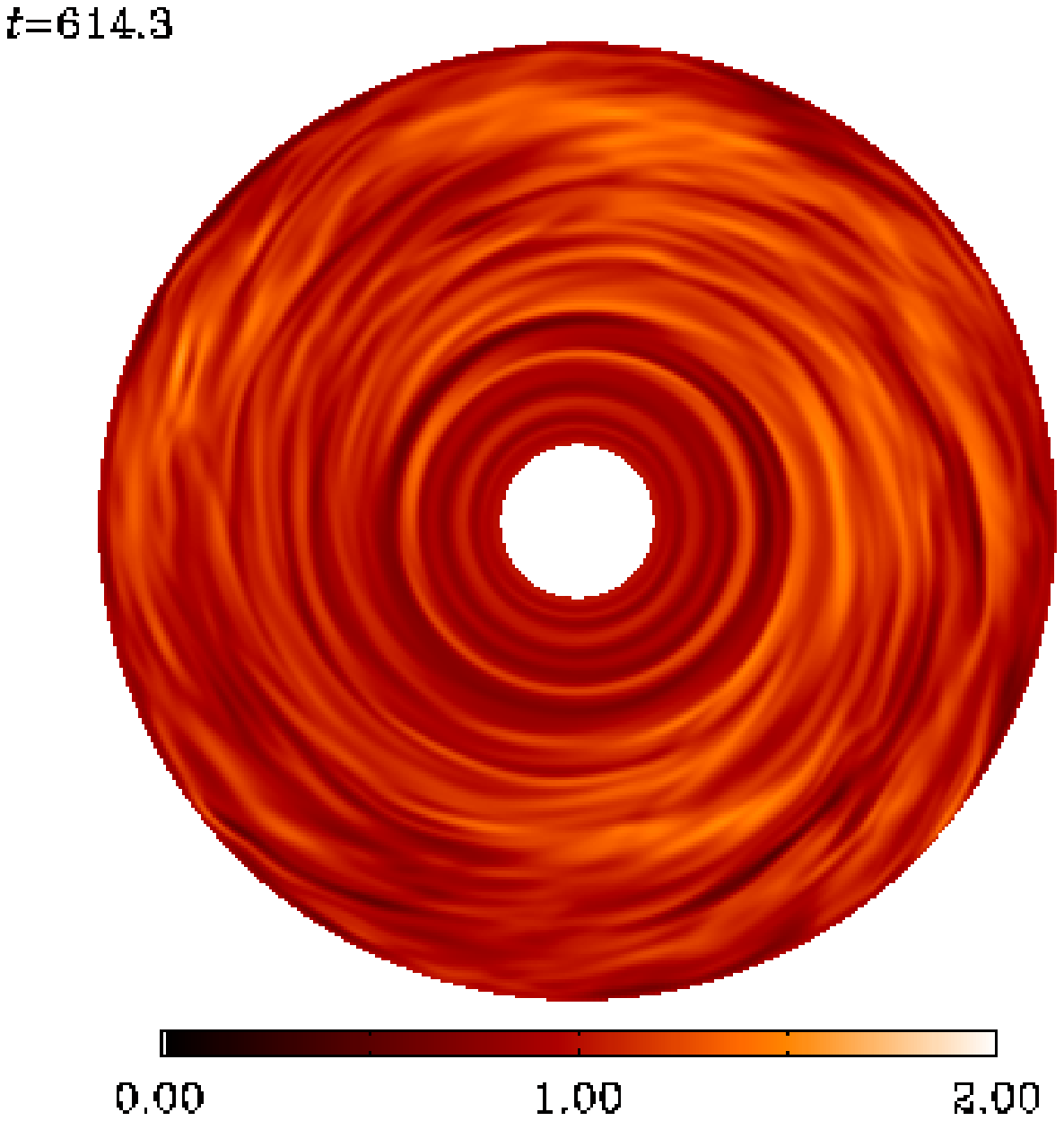}}
\resizebox{ 7cm}{!}{\includegraphics{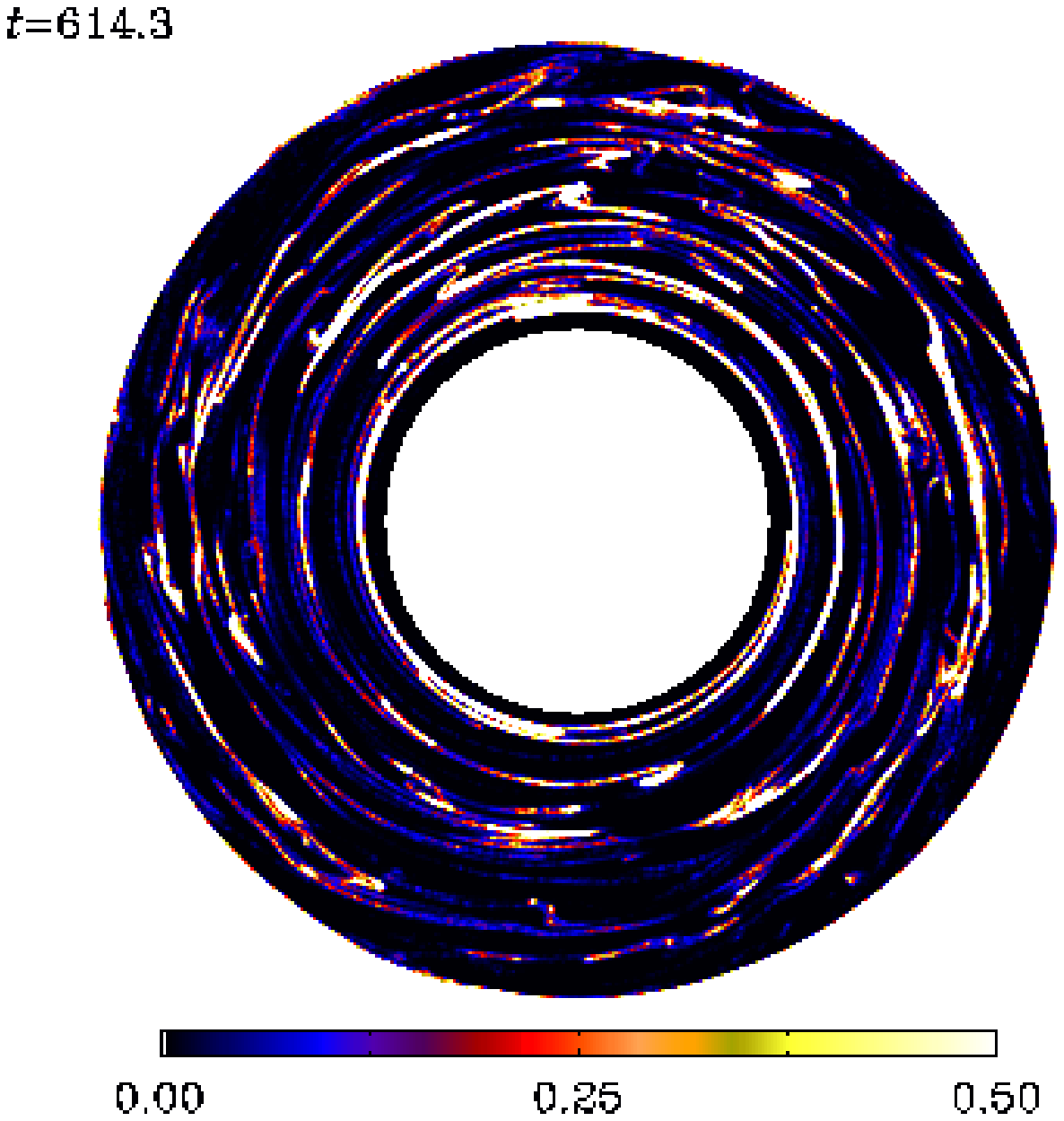}}
\resizebox{13cm}{!}{\includegraphics{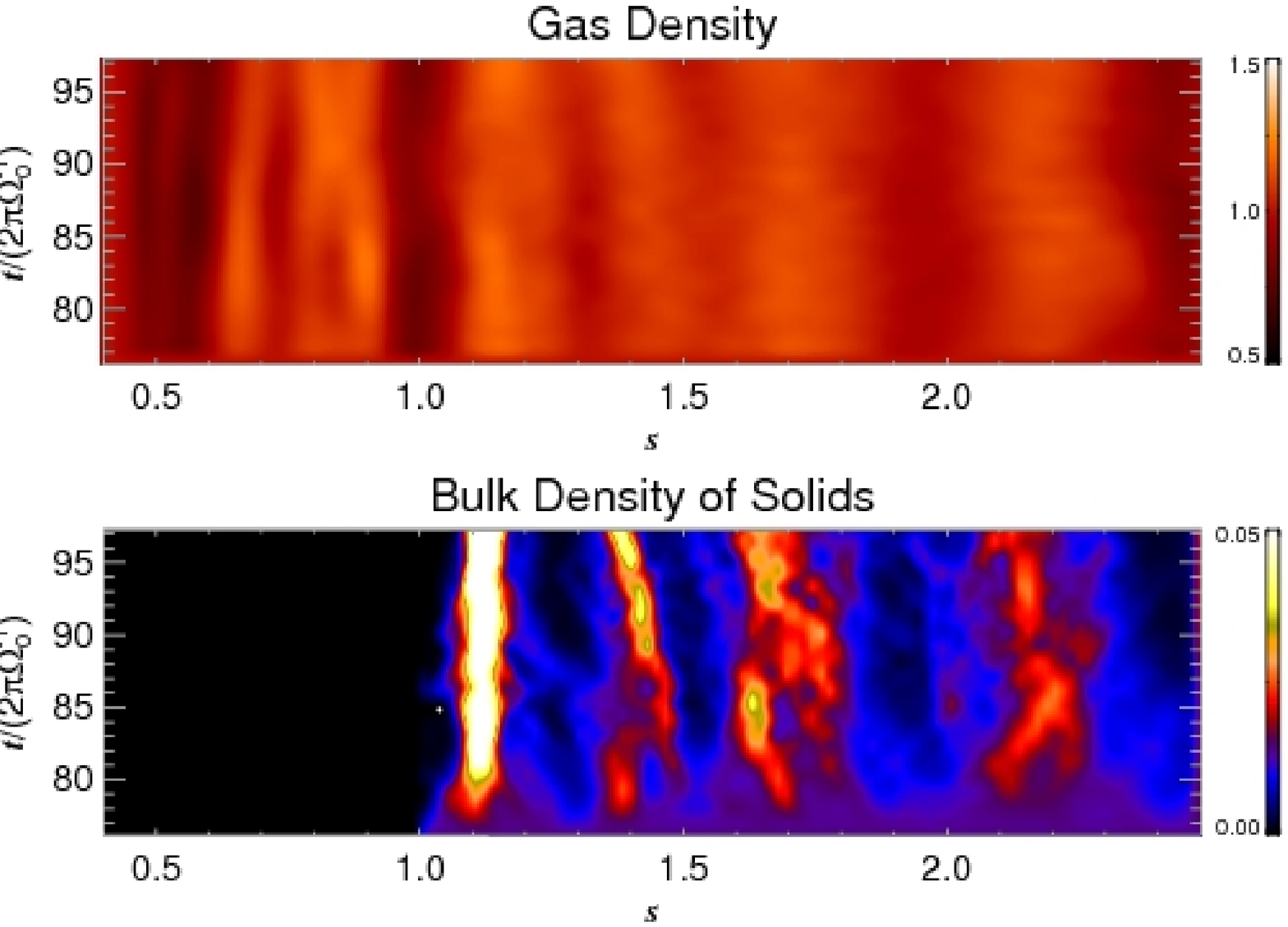}}
\end{center}
\caption[]{Density contours at midplane of the gas and solid phases of the disk
(left and right panel, respectively). The snapshots were taken at 20 orbits at
$s_0$ after the insertion of the particles. The color code for the solid phase
is selected to represent 2 sigma (0.25) above the average bulk density of 0.03.
The bright areas are saturated as the maximum density reaches as far as 85. A
movie of this simulation can be found at {\tt
http://www.astro.uu.se/$\sim$wlyra/planet.html}. A correlation with gas density
is seen, since the bright clumps of solids correspond to pressure maxima, i.e,
areas of high gas density.} 
\label{midplanes}
\end{figure*}

In Fig.~\ref{midplanes} we show contours of the gas density (left panel) and
the bulk density of solids (right panel) in a snapshot taken after 20 orbits
after the insertion of the particles.  A correlation is seen as the solids show
large concentration at areas of high gas density. Initially, the density
increases linearly as the particles sediment towards the midplane. As seen
before, the sedimentation is complete at the outer radius after $\approx$4
orbits. After that, the growth is only due to the particles being concentrated
in transient gas high pressures. 

\begin{figure}
\begin{center}
\resizebox{8cm}{!}{\includegraphics{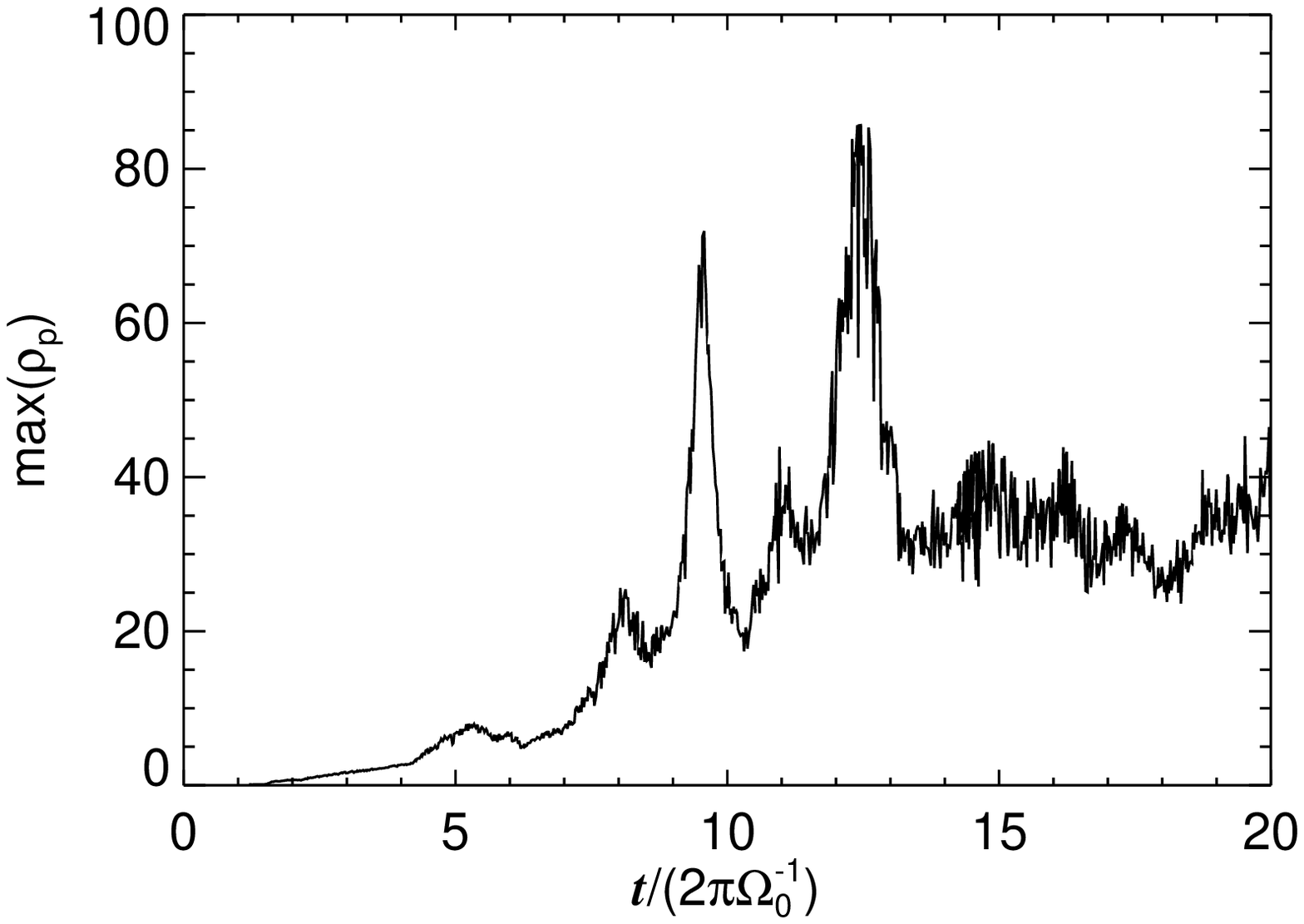}}
\end{center}
\caption[]{Maximum bulk density of solids, in units of the mean gas density as
a function of time for model A. Time is quoted in orbits at $s_0$. The maximum
density rises as the particles sediment towards the midplane. After the
sedimentation that lasts for four orbits, the particles are coupled with the
gas and trapping in transient gas high pressures raise their maximum density
well above average. The maximum density is usually around 30, but between
orbits 12 and 13, it reached values as high as 85.}
\label{maximum-density}
\end{figure}

The average bulk density is quite low ($\rho_{\rm p}$=0.003, or
$\xtimes{6.0}{-11}\,\rm{kg\,m^{-3}}$ in physical units, see Table~1), 
and several areas devoid of particles are seen in the disk.
However, the overdensities observed as bright clumps in the snapshot are
several standard deviations above average. By plotting the maximum solid
density (which is roughly the solids-to-gas ratio) throughout the simulation
(Fig.~\ref{maximum-density}), we see that its value is usually around 30, but
it can reach values as high as 85. Such a behavior was seen in shearing box
simulations by Johansen et al. (2006), who report local enhancements of the
solids-to-gas ratio by a factor of 100, also pointing that such concentrations
are gravitationally unstable, thus being able to collapse to form km-sized
bodies.

As a control, we also simulated the settling for a non-turbulent, purely
laminar, unperturbed disk. Without high pressure regions, the particles simply
sedimented towards the midplane forming a thin homogeneous layer of solids.

\section{Summary and conclusions}

We have considered MHD models of global Keplerian disks in Cartesian grids.
These disk-in-a-box models are able to develop and sustain MHD turbulence, in
good agreement with published results achieved with cylindrical codes and
shearing boxes. In this first article of the series, we investigated the
dependence of the MRI with disk scale height and the dynamics of solid boulders
in the global hydromagnetic turbulence. 

As a numerical solver we have used the {\sc Pencil Code}. This
finite-difference code solves the non-conservative form of the dynamical
equations using sixth-order spatial derivatives, achieving spatial resolution
that approaches that of spectral methods. The numerical scheme is stabilized by
using hyperdissipation and shock dissipation terms, which enter as free parameters in the
dynamical equations. The effect of hyperdissipation is to quench unstable modes
in the small scales of the grid, while affecting the large scale motion as
little as possible, whereas shock dissipation is invoked to smear out large
divergences in the flow field. We choose these parameters by performing
series of 2D gap opening simulations with a Jupiter mass planet and comparing
them with a higher
resolution calculation without explicit dissipative terms. 

We find evidence that the turbulence generated by the magnetorotational
instability grows with the thermal pressure. The turbulent stresses depend on
thermal pressure obeying a power law of $0.24\pm0.03$, compatible with the
value of $0.25$ found in shearing box calculations by Sano et al. (2004). We
extend this result to a global disk showing that the rise in pressure increases
the turbulent stresses, thus raising the angular momentum transport (and
therefore the mass accretion rate) although the alpha viscosity value drops. 

We also notice two curious effects. First, the dominance of the radial
component of the turbulent kinetic energy increases with temperature. The
percentage of the total kinetic energy stored in the radial component is 40\%
for the cold model A, and 60\% for the hotter model C. Second, the ratio of 
stresses $-M^{s\phi}/R^{s\phi}$ diminished with increasing temperature. It 
is 5 for model A, and just 1.3 for model C. The same is seen in the model 
without inner boundary, where the ratio is 6.5 for model A2 and very close 
to 1 for model C2. This effect is 
unexpected since it is believed that the shear parameter alone controls the 
ratio of stresses (Pessah et al. 2006, Ogilvie \& Pringle 1996). 
From the shearing box data of Sano et al. (2004) the stress ratio seems to 
be constant with temperature.

One explanation could be that, according to
Eq.~\ref{MHStatic-equilibrium}, the angular velocity is sub-Keplerian and the
increasing effects of pressure from the colder to the hotter models modifies
the shear. Quantitatively, however, one sees that the pressure correction is
too small to account for the decrease in the stress ratio and, more importantly,
would have the opposite effect. According to the linearized equations for the
evolution of the turbulent fluctuations, the Maxwell stress couples with shear
$q\Omega$, and the Reynolds stress couples with the large scale vorticity
$w=(2-q)\Omega$ (Balbus \& Hawley 1998), where $q=-\partial\ln\Omega/\partial
\ln r$ is the shear rate. The pressure-corrected angular velocity of the gas
can be approximated from Eq.\ (\ref{MHStatic-equilibrium}) as
\begin{equation}
  \Omega \simeq \Omega_{\rm K} (1-\eta) \, ,
\end{equation}
where $\eta= (1/2) (\partial \ln P/\partial \ln r) (H/s)^2 >0$ is a parameter
often used to parameterize the strength of the global pressure gradient (see
e.g.\ Nakagawa et al.\ 1986). Typical values of $\eta$ lie between
$0.001$ and $0.1$.

The reduction of both the angular frequency and shear rate should reduce
the Maxwell stress. Our simulations show the opposite, with the Maxwell stress
increasing as the pressure is raised. Regarding the stress ratio, reducing the
shear increases this quantity since the Reynolds stress falls faster than the
Maxwell stress due to the stabilizing effect of the growing vorticity
(Abramowicz et al., 1996). Once again, we see the opposite effect.

As most of the analysis of turbulent thin accretion disks have focused on
locally isothermal simulations using $c_s \approx 0.05$, changing the field
configuration while keeping the temperature constant, such behavior has been
largely overlooked. Although the disk temperatures considered in this case are
quite extreme for disks around T-Tauri stars, circumplanetary disks are thought
to be rather thick (Klahr \& Kley 2006) and therefore the evolution of the MRI
in such disks is expected to be more similar to the hotter cases considered in
this paper (models CEG) than the colder ones.

We investigated the effect of an inner boundary in the evolution and
outcome of the turbulence. By using a Cartesian grid, an inner boundary can be
discarded provided we smooth the gravitational potential to avoid a singularity
in the flow. Models without an inner boundary do not show the spurious build-up
of magnetic pressure and Reynolds stress seen in the models with boundaries,
while the global stresses and alpha viscosities are similar in the two cases.

In treating the solids, we make use of a large number of particles, which
allows us to effectively map the particles back into the grid as a density
field without using fluid approaches. We monitor the settling of the particles
toward the midplane and the formation of a sedimentary layer when the solids
are subject to gas drag and the gravity from the central object. The effective
diffusion provided by the turbulence prevents further settling of solids, in
accordance with the results of Johansen \& Klahr (2005). By having the global
disk perspective, we could measure the radial dependence of the diffusion scale
height of the solid component. The measured scale heights imply turbulent
vertical diffusion coefficients with globally averaged Schmidt numbers of
1.0$\pm$0.2 for model A ($\alpha\approx\ttimes{-3}$) and 0.78$\pm$0.06 for 
model D ($\alpha\approx\ttimes{-1}$).

We conclude that the models presented in this first paper of the series are
capable of sustaining turbulence and are adequately suited for further studies
of planet formation. Future papers will present studies of thermodynamics and
radiative transfer in the evolution of the turbulent stresses, planet-planet
and planet-disk interaction, the effect of stratification and the dynamics of
dead zones. 

\begin{acknowledgements}
The authors thank the referee, Dr. Ulf Torkelsson, for his many
constructive comments that helped improve the manuscript. We warmly 
acknowledge Dr. Paul Barklem for his help on improving the quality of the English. Simulations were performed at the PIA cluster of the Max-Planck-Institut
f{\"u}r Astronomie and on the Uppsala Multidisciplinary Center for Advanced
Computational Science (UPPMAX). 
\end{acknowledgements}

\begin{appendix}
\section{Anisotropic hyperdissipation}
\label{ch:hyperdissipation}

Hyperdissipation is used to quench unstable modes at the grid scale, therefore
being intrinsically resolution-dependent. Because of this, {\em isotropic}
dissipation only gives equal dissipation in all spatial directions if $\Delta
x=\Delta y=\Delta z$, i.e., if the cells are cubic. For non-cubic cells,
anisotropic dissipation is required as different directions may be better/worse
sampled, thus needing less/more numerical smoothing. Such a generalization is
straightforward. We notice that hyperdiffusion works as a conservative term in the continuity equation such that 
\begin{equation}
  \label{apcontinuity}
  f_D(\rho) = \del \cdot \vv{{\mathcal J}},
\end{equation}
where $\vv{{\mathcal J}}=D_3 \del^5 \rho$ is the mass flux 
due to hyperdiffusion. For simplicity, we will drop the 
subscripts ``3'' from the coefficients hereafter. This formulation 
reduces to the usual sixth-order hyperdiffusion under the condition that $D$ is 
constant. Generalizing it to three dimensions simply involves replacing this 
mass flow by
\[
  \vv{{\mathcal J}}=\left(D_x \pderivn{\rho}{x}{5}, D_y \pderivn{\rho}{y}{5}, D_z \pderivn{\rho}{z}{5}\right),
\]
so that different diffusion operates in different directions. Since $D_x$,
$D_y$ and $D_z$ are constants, the divergent of this vector is 
\[
  \Div\vv{{\mathcal J}}=D_x \pderivn{\rho}{x}{6}+ 
  D_y \pderivn{\rho}{y}{6}+
  D_z \pderivn{\rho}{z}{6}.
\]
The formulation for resistivity is strictly the same. For viscosity it also
assumes the same form if we consider a simple $n$th order rate of strain tensor
operator $S^{(n)}_{ij} = \partial{}^n_j{u_i}$.
 
\section{Shocks}
\label{ch:shocks}

Shock viscosity is taken to be proportional to positive flow convergence,
maximum over three zones, and smoothed to second order,
\begin{equation}
\label{shock-visc}
  \zeta_\nu=\nu_{\rm sh}\left<\max_3[(-\Div\vv{u})_+]\right>{\left[\min(\Delta x,\
\Delta y,\Delta z)\right]}^2,
\end{equation}
where $\nu_{\rm sh}$ is a constant defining the strength of the shock
viscosity, usually around unity. We refer to it as the shock viscosity
coefficient (Haugen et al. 2004). In the equation of motion it takes the form
of a bulk viscosity so that now the stress tensor contains
\begin{equation}
\tau_{ij}=[...]+\rho\zeta_\nu\delta_{ij}\Div\vv{u},
\end{equation}
where [...] refers to the (hyper) viscous terms described in
\App{ch:hyperdissipation}. The acceleration due to shock viscosity is therefore
\begin{eqnarray}
f_\nu(\vv{u},\rho) &=& \rho^{-1}\Div\tau \nonumber \\
&=&\zeta_{\nu}\left[\del(\Div\vv{u})+\left(\del\ln\rho+\del\ln\zeta_{\nu}\right)\Div\vv{u}\right].  \nonumber \\
&&
\end{eqnarray}
Such a viscosity scheme ensures that energy is dissipated in regions of the
flow where shocks occur, whereas more quiescent regions are left untouched. The
formulations for shock diffusion and shock resistivity are similar, yielding
\begin{equation}
f_D(\rho)= \zeta_D\left(\Laplace\rho+\del\ln\zeta_D\cdot\del\rho\right),
\end{equation}
and
\begin{equation}
f_\eta(\vv{A})= \zeta_\eta\left(\Laplace\vv{A}+\del\ln\zeta_\eta\Div\vv{A}\right),
\end{equation}
where $\zeta_D$ and $\zeta_\eta$ are analogous to $\zeta_\nu$ in
Eq.~(\ref{shock-visc}), containing their respective shock diffusion and shock
resistivity coefficients $D_{\rm sh}$ and $\eta_{\rm sh}$.

\end{appendix}

\end{document}